\documentclass[final]{IEEEtran}
\usepackage[update,prepend]{epstopdf}
\usepackage{amsmath}
\usepackage{algorithm}

\usepackage{array}
\usepackage{mathrsfs}
\usepackage{graphics}
\usepackage{multirow}
\usepackage{tikz}
\usepackage{bbm} % for \mathbbm{1}
\usepackage{pdfpages}
\usepackage{multirow}
\usepackage{subfig}
\usepackage{comment}
\usepackage{cite}
\usepackage{makecell}

\usepackage{setspace}	% Remove in double column version. Also search for \setstretch in the body of the paper and comment these commands for double column
\usepackage{graphicx}
\usepackage{algorithm,algorithmic}
\usepackage{multicol}

\usepackage[justification=centering]{caption}
\usepackage{textcomp}
\usepackage{psfrag}
\usepackage{arydshln}
\usepackage{url}
\usepackage{soul}
\usepackage{graphicx,color}
\usepackage[nolist]{acronym}
\usepackage{array}

\usepackage{mathtools,lipsum}
\usepackage{cuted}
\usepackage{amsmath}
\usepackage{graphicx}
\usepackage{threeparttable}

\def\nb0{{\mathbf{0}}}
\def\nb1{{\mathbf{1}}}

\begin{document}

%\pagenumbering{gobble}
\graphicspath{{./Figures/}}
	\begin{acronym}

\acro{5G-NR}{5G New Radio}
\acro{3GPP}{3rd Generation Partnership Project}
\acro{ABS}{aerial base station}
\acro{AC}{address coding}
\acro{ACF}{autocorrelation function}
\acro{ACR}{autocorrelation receiver}
\acro{ADC}{analog-to-digital converter}
\acrodef{aic}[AIC]{Analog-to-Information Converter}     
\acro{AIC}[AIC]{Akaike information criterion}
\acro{aric}[ARIC]{asymmetric restricted isometry constant}
\acro{arip}[ARIP]{asymmetric restricted isometry property}

\acro{ARQ}{Automatic Repeat Request}
\acro{AUB}{asymptotic union bound}
\acrodef{awgn}[AWGN]{Additive White Gaussian Noise}     
\acro{AWGN}{additive white Gaussian noise}

\acro{APSK}[PSK]{asymmetric PSK} 

\acro{waric}[AWRICs]{asymmetric weak restricted isometry constants}
\acro{warip}[AWRIP]{asymmetric weak restricted isometry property}
\acro{BCH}{Bose, Chaudhuri, and Hocquenghem}        
\acro{BCHC}[BCHSC]{BCH based source coding}
\acro{BEP}{bit error probability}
\acro{BFC}{block fading channel}
\acro{BG}[BG]{Bernoulli-Gaussian}
\acro{BGG}{Bernoulli-Generalized Gaussian}
\acro{BPAM}{binary pulse amplitude modulation}
\acro{BPDN}{Basis Pursuit Denoising}
\acro{BPPM}{binary pulse position modulation}
\acro{BPSK}{Binary Phase Shift Keying}
\acro{BPZF}{bandpass zonal filter}
\acro{BSC}{binary symmetric channels}              
\acro{BU}[BU]{Bernoulli-uniform}
\acro{BER}{bit error rate}
\acro{BS}{base station}
\acro{BW}{BandWidth}
\acro{BLLL}{ binary log-linear learning }

\acro{CP}{Cyclic Prefix}
\acrodef{cdf}[CDF]{cumulative distribution function}   
\acro{CDF}{Cumulative Distribution Function}
\acrodef{c.d.f.}[CDF]{cumulative distribution function}
\acro{CCDF}{complementary cumulative distribution function}
\acrodef{ccdf}[CCDF]{complementary CDF}               
\acrodef{c.c.d.f.}[CCDF]{complementary cumulative distribution function}
\acro{CD}{cooperative diversity}

\acro{CDMA}{Code Division Multiple Access}
\acro{ch.f.}{characteristic function}
\acro{CIR}{channel impulse response}
\acro{cosamp}[CoSaMP]{compressive sampling matching pursuit}
\acro{CR}{cognitive radio}
\acro{cs}[CS]{compressed sensing}                   
\acrodef{cscapital}[CS]{Compressed sensing} %will not include it in the list
\acrodef{CS}[CS]{compressed sensing}
\acro{CSI}{channel state information}
\acro{CCSDS}{consultative committee for space data systems}
\acro{CC}{convolutional coding}
\acro{Covid19}[COVID-19]{Coronavirus disease}

\acro{DAA}{detect and avoid}
\acro{DAB}{digital audio broadcasting}
\acro{DCT}{discrete cosine transform}
\acro{dft}[DFT]{discrete Fourier transform}
\acro{DR}{distortion-rate}
\acro{DS}{direct sequence}
\acro{DS-SS}{direct-sequence spread-spectrum}
\acro{DTR}{differential transmitted-reference}
\acro{DVB-H}{digital video broadcasting\,--\,handheld}
\acro{DVB-T}{digital video broadcasting\,--\,terrestrial}
\acro{DL}{DownLink}
\acro{DSSS}{Direct Sequence Spread Spectrum}
\acro{DFT-s-OFDM}{Discrete Fourier Transform-spread-Orthogonal Frequency Division Multiplexing}
\acro{DAS}{Distributed Antenna System}
\acro{DNA}{DeoxyriboNucleic Acid}

\acro{EC}{European Commission}
\acro{EED}[EED]{exact eigenvalues distribution}
\acro{EIRP}{Equivalent Isotropically Radiated Power}
\acro{ELP}{equivalent low-pass}
\acro{eMBB}{Enhanced Mobile Broadband}
\acro{EMF}{ElectroMagnetic Field}
\acro{EU}{European union}
\acro{EI}{Exposure Index}
\acro{eICIC}{enhanced Inter-Cell Interference Coordination}

\acro{FC}[FC]{fusion center}
\acro{FCC}{Federal Communications Commission}
\acro{FEC}{forward error correction}
\acro{FFT}{fast Fourier transform}
\acro{FH}{frequency-hopping}
\acro{FH-SS}{frequency-hopping spread-spectrum}
\acrodef{FS}{Frame synchronization}
\acro{FSsmall}[FS]{frame synchronization}  
\acro{FDMA}{Frequency Division Multiple Access}

\acro{GA}{Gaussian approximation}
\acro{GF}{Galois field }
\acro{GG}{Generalized-Gaussian}
\acro{GIC}[GIC]{generalized information criterion}
\acro{GLRT}{generalized likelihood ratio test}
\acro{GPS}{Global Positioning System}
\acro{GMSK}{Gaussian Minimum Shift Keying}
\acro{GSMA}{Global System for Mobile communications Association}
\acro{GS}{ground station}
\acro{GMG}{ Grid-connected MicroGeneration}

\acro{HAP}{high altitude platform}
\acro{HetNet}{Heterogeneous network}

\acro{IDR}{information distortion-rate}
\acro{IFFT}{inverse fast Fourier transform}
\acro{iht}[IHT]{iterative hard thresholding}
\acro{i.i.d.}{independent, identically distributed}
\acro{IoT}{Internet of Things}                      
\acro{IR}{impulse radio}
\acro{lric}[LRIC]{lower restricted isometry constant}
\acro{lrict}[LRICt]{lower restricted isometry constant threshold}
\acro{ISI}{intersymbol interference}
\acro{ITU}{International Telecommunication Union}
\acro{ICNIRP}{International Commission on Non-Ionizing Radiation Protection}
\acro{IEEE}{Institute of Electrical and Electronics Engineers}
\acro{ICES}{IEEE international committee on electromagnetic safety}
\acro{IEC}{International Electrotechnical Commission}
\acro{IARC}{International Agency on Research on Cancer}
\acro{IS-95}{Interim Standard 95}

\acro{KPI}{Key Performance Indicator}

\acro{LEO}{low earth orbit}
\acro{LF}{likelihood function}
\acro{LLF}{log-likelihood function}
\acro{LLR}{log-likelihood ratio}
\acro{LLRT}{log-likelihood ratio test}
\acro{LoS}{Line-of-Sight}
\acro{LRT}{likelihood ratio test}
\acro{wlric}[LWRIC]{lower weak restricted isometry constant}
\acro{wlrict}[LWRICt]{LWRIC threshold}
\acro{LPWAN}{Low Power Wide Area Network}
\acro{LoRaWAN}{Low power long Range Wide Area Network}
\acro{NLoS}{Non-Line-of-Sight}
\acro{LiFi}[Li-Fi]{light-fidelity}
 \acro{LED}{light emitting diode}
 \acro{LABS}{LoS transmission with each ABS}
 \acro{NLABS}{NLoS transmission with each ABS}

\acro{MB}{multiband}
\acro{MC}{macro cell}
\acro{MDS}{mixed distributed source}
\acro{MF}{matched filter}
\acro{m.g.f.}{moment generating function}
\acro{MI}{mutual information}
\acro{MIMO}{Multiple-Input Multiple-Output}
\acro{MISO}{multiple-input single-output}
\acrodef{maxs}[MJSO]{maximum joint support cardinality}                       
\acro{ML}[ML]{maximum likelihood}
\acro{MMSE}{minimum mean-square error}
\acro{MMV}{multiple measurement vectors}
\acrodef{MOS}{model order selection}
\acro{M-PSK}[${M}$-PSK]{$M$-ary phase shift keying}                       
\acro{M-APSK}[${M}$-PSK]{$M$-ary asymmetric PSK} 
\acro{MP}{ multi-period}
\acro{MINLP}{mixed integer non-linear programming}

\acro{M-QAM}[$M$-QAM]{$M$-ary quadrature amplitude modulation}
\acro{MRC}{maximal ratio combiner}                  
\acro{maxs}[MSO]{maximum sparsity order}                                      
\acro{M2M}{Machine-to-Machine}                                                
\acro{MUI}{multi-user interference}
\acro{mMTC}{massive Machine Type Communications}      
\acro{mm-Wave}{millimeter-wave}
\acro{MP}{mobile phone}
\acro{MPE}{maximum permissible exposure}
\acro{MAC}{media access control}
\acro{NB}{narrowband}
\acro{NBI}{narrowband interference}
\acro{NLA}{nonlinear sparse approximation}
\acro{NLOS}{Non-Line of Sight}
\acro{NTIA}{National Telecommunications and Information Administration}
\acro{NTP}{National Toxicology Program}
\acro{NHS}{National Health Service}

\acro{LOS}{Line of Sight}

\acro{OC}{optimum combining}                             
\acro{OC}{optimum combining}
\acro{ODE}{operational distortion-energy}
\acro{ODR}{operational distortion-rate}
\acro{OFDM}{Orthogonal Frequency-Division Multiplexing}
\acro{omp}[OMP]{orthogonal matching pursuit}
\acro{OSMP}[OSMP]{orthogonal subspace matching pursuit}
\acro{OQAM}{offset quadrature amplitude modulation}
\acro{OQPSK}{offset QPSK}
\acro{OFDMA}{Orthogonal Frequency-division Multiple Access}
\acro{OPEX}{Operating Expenditures}
\acro{OQPSK/PM}{OQPSK with phase modulation}

\acro{PAM}{pulse amplitude modulation}
\acro{PAR}{peak-to-average ratio}
\acrodef{pdf}[PDF]{probability density function}                      
\acro{PDF}{probability density function}
\acrodef{p.d.f.}[PDF]{probability distribution function}
\acro{PDP}{power dispersion profile}
\acro{PMF}{probability mass function}                             
\acrodef{p.m.f.}[PMF]{probability mass function}
\acro{PN}{pseudo-noise}
\acro{PPM}{pulse position modulation}
\acro{PRake}{Partial Rake}
\acro{PSD}{power spectral density}
\acro{PSEP}{pairwise synchronization error probability}
\acro{PSK}{phase shift keying}
\acro{PD}{power density}
\acro{8-PSK}[$8$-PSK]{$8$-phase shift keying}
\acro{PPP}{Poisson point process}
\acro{PCP}{Poisson cluster process}
 
\acro{FSK}{Frequency Shift Keying}

\acro{QAM}{Quadrature Amplitude Modulation}
\acro{QPSK}{Quadrature Phase Shift Keying}
\acro{OQPSK/PM}{OQPSK with phase modulator }

\acro{RD}[RD]{raw data}
\acro{RDL}{"random data limit"}
\acro{ric}[RIC]{restricted isometry constant}
\acro{rict}[RICt]{restricted isometry constant threshold}
\acro{rip}[RIP]{restricted isometry property}
\acro{ROC}{receiver operating characteristic}
\acro{rq}[RQ]{Raleigh quotient}
\acro{RS}[RS]{Reed-Solomon}
\acro{RSC}[RSSC]{RS based source coding}
\acro{r.v.}{random variable}                               
\acro{R.V.}{random vector}
\acro{RMS}{root mean square}
\acro{RFR}{radiofrequency radiation}
\acro{RIS}{Reconfigurable Intelligent Surface}
\acro{RNA}{RiboNucleic Acid}
\acro{RRM}{Radio Resource Management}
\acro{RUE}{reference user equipments}
\acro{RAT}{radio access technology}
\acro{RB}{resource block}

\acro{SA}[SA-Music]{subspace-augmented MUSIC with OSMP}
\acro{SC}{small cell}
\acro{SCBSES}[SCBSES]{Source Compression Based Syndrome Encoding Scheme}
\acro{SCM}{sample covariance matrix}
\acro{SEP}{symbol error probability}
\acro{SG}[SG]{sparse-land Gaussian model}
\acro{SIMO}{single-input multiple-output}
\acro{SINR}{signal-to-interference plus noise ratio}
\acro{SIR}{signal-to-interference ratio}
\acro{SISO}{Single-Input Single-Output}
\acro{SMV}{single measurement vector}
\acro{SNR}[\textrm{SNR}]{signal-to-noise ratio} 
\acro{sp}[SP]{subspace pursuit}
\acro{SS}{spread spectrum}
\acro{SW}{sync word}
\acro{SAR}{specific absorption rate}
\acro{SSB}{synchronization signal block}
\acro{SR}{shrink and realign}

\acro{tUAV}{tethered Unmanned Aerial Vehicle}
\acro{TBS}{terrestrial base station}

\acro{uUAV}{untethered Unmanned Aerial Vehicle}
\acro{PDF}{probability density functions}

\acro{PL}{path-loss}

\acro{TH}{time-hopping}
\acro{ToA}{time-of-arrival}
\acro{TR}{transmitted-reference}
\acro{TW}{Tracy-Widom}
\acro{TWDT}{TW Distribution Tail}
\acro{TCM}{trellis coded modulation}
\acro{TDD}{Time-Division Duplexing}
\acro{TDMA}{Time Division Multiple Access}
\acro{Tx}{average transmit}

\acro{UAV}{Unmanned Aerial Vehicle}
\acro{uric}[URIC]{upper restricted isometry constant}
\acro{urict}[URICt]{upper restricted isometry constant threshold}
\acro{UWB}{ultrawide band}
\acro{UWBcap}[UWB]{Ultrawide band}   
\acro{URLLC}{Ultra Reliable Low Latency Communications}
         
\acro{wuric}[UWRIC]{upper weak restricted isometry constant}
\acro{wurict}[UWRICt]{UWRIC threshold}                
\acro{UE}{User Equipment}
\acro{UL}{UpLink}

\acro{WiM}[WiM]{weigh-in-motion}
\acro{WLAN}{wireless local area network}
\acro{wm}[WM]{Wishart matrix}                               
\acroplural{wm}[WM]{Wishart matrices}
\acro{WMAN}{wireless metropolitan area network}
\acro{WPAN}{wireless personal area network}
\acro{wric}[WRIC]{weak restricted isometry constant}
\acro{wrict}[WRICt]{weak restricted isometry constant thresholds}
\acro{wrip}[WRIP]{weak restricted isometry property}
\acro{WSN}{wireless sensor network}                        
\acro{WSS}{Wide-Sense Stationary}
\acro{WHO}{World Health Organization}
\acro{Wi-Fi}{Wireless Fidelity}

\acro{sss}[SpaSoSEnc]{sparse source syndrome encoding}

\acro{VLC}{Visible Light Communication}
\acro{VPN}{Virtual Private Network} 
\acro{RF}{Radio Frequency}
\acro{FSO}{Free Space Optics}
\acro{IoST}{Internet of Space Things}

\acro{GSM}{Global System for Mobile Communications}
\acro{2G}{Second-generation cellular network}
\acro{3G}{Third-generation cellular network}
\acro{4G}{Fourth-generation cellular network}
\acro{5G}{Fifth-generation cellular network}	
\acro{gNB}{next-generation Node-B Base Station}
\acro{NR}{New Radio}
\acro{UMTS}{Universal Mobile Telecommunications Service}
\acro{LTE}{Long Term Evolution}

\acro{QoS}{Quality of Service}
\end{acronym}
	
	%% EMF definitions
\newcommand{\SAR} {\mathrm{SAR}}
\newcommand{\WBSAR} {\mathrm{SAR}_{\mathsf{WB}}}
\newcommand{\gSAR} {\mathrm{SAR}_{10\si{\gram}}}
\newcommand{\Sab} {S_{\mathsf{ab}}}
\newcommand{\Eavg} {E_{\mathsf{avg}}}
\newcommand{\ft}{f_{\textsf{th}}}
\newcommand{\alphatf}{\alpha_{24}}

\title{
Modeling and Analysis of Non-Terrestrial Networks by Spherical Stochastic Geometry
}
\author{
Ruibo Wang, {\em Member, IEEE}, Mustafa A. Kishk, {\em Member, IEEE}, and \\ Mohamed-Slim Alouini, {\em Fellow, IEEE}
\thanks{Ruibo Wang and Mohamed-Slim Alouini are with King Abdullah University of Science and Technology (KAUST), CEMSE division, Thuwal 23955-6900, Saudi Arabia. Mustafa A. Kishk is with the Department of Electronic Engineering, Maynooth University, Maynooth, W23 F2H6, Ireland. (e-mail: ruibo.wang@kaust.edu.sa; mustafa.kishk@mu.ie; slim.alouini@kaust.edu.sa). 
}
\vspace{-6mm}
}
\maketitle
\thispagestyle{empty}

\begin{abstract}
Non-terrestrial networks (NTNs) are anticipated to be indispensable in extending coverage and enabling global communication access in next-generation wireless networks. With the extensive deployment of non-terrestrial platforms, evaluating the performance of NTN-enabled communication systems becomes a challenging task. Spherical stochastic geometry (SG) is a recently proposed analytical framework that has garnered increasing attention. Due to its suitability for modeling large-scale dynamic topologies and its ability to provide an analytical framework for interference analysis and low-complexity performance evaluation, spherical SG has been widely applied in NTN performance analysis. This paper surveys the modeling and analysis of NTN networks based on spherical SG. We begin by introducing the spherical SG framework, detailing its history and development. Next, we categorize existing spherical SG models into three types based on orbital modeling methods and provide algorithm implementations for common models. Furthermore, we investigate the accuracy and necessity of spherical modeling through case studies. On the topology level, concepts such as association strategy, central angle, zenith angle, contact angle, and availability probability are introduced, with simple derivations provided. On the channel level, we detail the modeling of large-scale fading, small-scale fading, and beam gain for different channel links. Finally, we discuss several advanced topics that have not been fully explored but have strong motivation and research potential, and we predict future research directions.
\end{abstract}

\section{Introduction}\label{section1}
In 5G and beyond communication systems, non-terrestrial networks (NTNs) play an irreplaceable role in enhancing communication coverage, ensuring service continuity, and guaranteeing seamless emergency communications \cite{rinaldi2020non}. NTNs can cover areas that are difficult for terrestrial networks to reach, such as remote regions, oceans, and mountains, extending communication services to every corner of the globe \cite{giordani2020non}. Additionally, in the event of terrestrial network disruptions caused by natural disasters, NTNs can quickly provide alternative communication services, ensuring seamless emergency communications and improving the efficiency of disaster response and rescue operations \cite{azari2022evolution}. Furthermore, NTNs can serve as backup systems for terrestrial networks, enhancing the reliability and stability of communication networks. In the event of terrestrial network failures, NTNs can take over part of the communication traffic, reducing the impact of service interruptions \cite{belmekki2022unleashing}. 

\par
\begin{figure*}[t]
\centering
\vspace{-0.2cm}
\includegraphics[width = 0.9\textwidth]{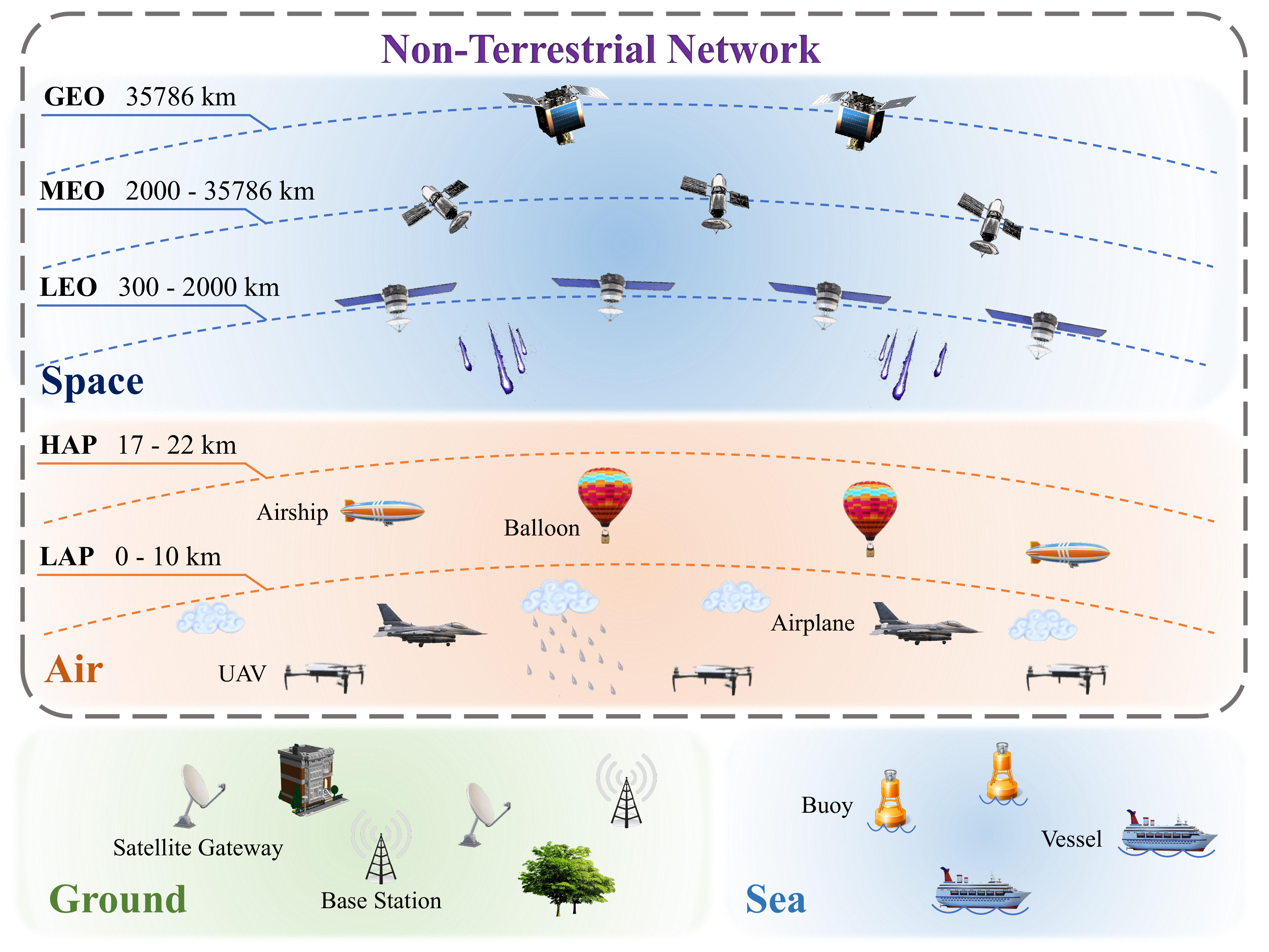}
\caption{Illustration of a space-air-ground-sea integrated network.}
\vspace{-0.2cm}
\label{figure1-1}
\end{figure*}

As shown in Fig.~\ref{figure1-1}, NTN consists of low-altitude platforms (LAPs), high-altitude platforms (HAPs), and satellites, all of which are collectively referred to as non-terrestrial platforms (NTPs). Since NTN involves multiple different network tiers, performance analysis is particularly important. NTPs operate at various altitudes and environments, which have different climate and electromagnetic interference conditions \cite{liu2018space}. These variations can affect signal propagation characteristics, thus impacting network performance and presenting challenges for performance analysis. Additionally, NTPs often move at high speeds or operate in orbit. Orbital changes and flight paths lead to frequent changes in network topology, increasing the complexity of performance evaluation. 

\par
In recent years, the development of spherical stochastic geometry (SG) analytical frameworks has provided new perspectives and tools for NTN performance analysis. In the following subsections of the introduction, the topological features of NTPs are first introduced, serving as important references for SG-based modeling. Next, we explain the framework and core concepts of spherical SG, its history and development, and the motivations for applying spherical SG to NTN performance analysis. Finally, we outline the differences between this survey and other tutorials on SG, and further detail the organization and contributions of this survey.

\begin{table}[ht]
\centering
\caption{List of abbreviations.}
\label{tableI-2}
\renewcommand{\arraystretch}{1.1}
\begin{tabular}{|c|c|}
\hline
Acronyms  &  Description  \\ \hline \hline
AN & Artificial noise \\ \hline
BPP  & Binomial point process \\ \hline
CPP  & Cox point process \\ \hline
CRLB & Cram$\acute{e}$r–Rao lower
bound \\ \hline
DSBPP & Dual-stochastic binomial point process  \\ \hline
GEO & Geostationary Earth orbit \\ \hline
GNSS & Global navigation satellite system \\ \hline
GPS & Global positioning system \\ \hline
HAP & High altitude platform \\ \hline
IoT & Internet of things \\ \hline
ISL & Inter-satellite link \\ \hline
LAP & Low altitude platform \\ \hline
LEO & Low Earth orbit \\ \hline
LoS & Line-of-sight \\ \hline
MEO & Medium Earth orbit \\ \hline
NLoS & Non-line-of-sight \\ \hline
NTN & Non-terrestrial network \\ \hline
NTP & Non-terrestrial platform \\ \hline
OGM & Orbit geometry model \\ \hline
PCP & Poisson cluster process \\ \hline
PDF & Probability density function \\ \hline 
PGFL & Probability generating functional \\ \hline
PLP & Possion line process \\ \hline
PLS & Physical layer security \\ \hline
PPP  & Poisson point process \\ \hline
SAGIN & Space-air-ground integrated network \\ \hline
SAGSIN & Space-air-ground-sea integrated network \\ \hline
SG  & Stochastic geometry \\ \hline 
SINR & Signal-to-interference plus noise ratio \\ \hline
SR & Shadowed Rician \\ \hline
STK & Systems tool kit \\ \hline
TDoA & Time difference of arrival \\ \hline
UAV & Unmanned aerial vehicle \\ \hline
WCS & Wireless charging stations \\ \hline
\end{tabular}
\end{table}

\subsection{Topological Features of NTPs}
This subsection describes the topological features of NTPs mentioned in existing spherical SG research, which will serve as the basis for spherical SG modeling and the foundation for topological analysis within the SG framework.

\subsubsection{LEO Satellite}
Low Earth orbit (LEO) satellites refer to satellites that operate at an altitude of $300$ to $2000$~km above the Earth surface. Most research in the field of spherical SG focuses on LEO satellites as the research object. LEO satellites orbit in a fixed trajectory in a periodic manner. Due to their lower orbit, LEO satellites complete an orbit around the Earth in a relatively short time, typically between $90$ and $120$ minutes. This means they move quickly relative to the Earth. The Walker constellation is one of the most common configurations for LEO satellite constellations \cite{walker1984satellite}. In the Walker constellation, the satellite orbits are circular, with the orbital planes evenly distributed, and the satellites within each orbital plane are uniformly spaced. Based on whether the satellite orbits pass or nearly pass through the Earth's poles, LEO constellations can be classified into polar orbit constellations and non-polar orbit constellations. For polar orbit constellations, all orbits converge at the poles, so the density of satellites at the poles will be significantly higher than at the equator. As shown in Fig.~\ref{figure2-1}, the Iridium constellation \cite{Iridium} and the Starlink constellation \cite{korobkov2020traffic} are representative polar orbit constellation and non-polar orbit constellations, respectively.

\subsubsection{MEO Satellite}
Medium Earth orbit (MEO) satellites are located at altitudes ranging from $2000$ to $35786$~km above the Earth surface. Compared to LEO satellites, a single MEO satellite covers a larger area. Therefore, MEO satellite constellations tend to be smaller in scale, usually consisting of a few dozen satellites. Currently, MEO satellites are mainly used for global navigation satellite systems such as the global positioning system (GPS), Beidou, and Galileo \cite{cai2015precise}. Additionally, the constellation configurations of MEO and LEO satellites are similar. 

\subsubsection{GEO Satellite}
Geostationary Earth Orbit (GEO) satellites are located at an altitude of $35786$~km above the Earth surface. They rotate at the same speed as the Earth's rotation, making them geostationary. Each GEO satellite can cover about one-third of the Earth's surface, so three GEO satellites can provide global coverage. According to the two-line element dataset given in \cite{piergentili2014close}, there are currently $391$ GEO satellites deployed in circular orbits above the Earth's equator.

\subsubsection{HAP}
{\color{black}
HAP is a type of aerial platform at an altitude of around $20$~km, such as balloons and airships. According to their trajectories, HAP can be classified into stationary trajectories, circular trajectories, and cyclical trajectories. While HAP can be designed to move within specific regions, they typically do not orbit the Earth in fixed paths like satellites, and their range of movement is much smaller compared to satellites. Within the spherical SG framework, HAP provides downlink coverage to ground users or serves as a communication relay in the space-air-ground integrated network (SAGIN). }

\subsubsection{Aircraft}
In the context of a spherical SG framework, the aircraft being studied are primarily LAPs that fly along fixed orbits at altitudes ranging from $1$ to $10$~km above ground. They mainly rely on a space-to-air link and connect to the network via satellites. Unlike global modeling of HAPs and satellite constellations, the locations of aircraft are limited to fixed trajectories. Lower-altitude unmanned aerial vehicles (UAVs) are modeled as a planar point process and are therefore not within the scope of this survey.

\subsection{Spherical SG Analytical Framework}
In this subsection, we introduce the SG analytical framework, the development of the spherical SG field, and the motivation and advantages of applying the spherical SG framework to NTNs. 

\subsubsection{Introduction of the SG Framework}
In the field of wireless communications, SG is a method that uses tools from probability theory and geometry to model and analyze wireless networks \cite{andrews2016primer}. It treats the positions of nodes (such as NTPs) in the network as a random point process, thereby providing a more flexible way to study the performance of wireless communication systems \cite{haenggi2009stochastic}. The core idea of SG is to model the positions of network nodes according to a specific point process in exchange for the analytical tractability of network performance \cite{zappone2019wireless}. For a given type of point process, if the characteristics and handling methods are diverse, the mathematical expressions for analyzing its topological structure are relatively simple, and the computational complexity for analytically deriving network performance information under this point process is low, then the point process is said to have strong tractability.

\par
The SG analytical framework is divided into three parts: modeling, topology analysis, and performance metrics analysis \cite{alzenad2019coverage}. Modeling includes modeling NTPs' spatial distribution and modeling the communication channel from NTP to users. Based on the spatial distribution model, we can analyze communication links and the network at the topology level. For example, we can derive the distance distribution from NTP to users and the probability of having available NTPs. Finally, based on the channel model and the topology-level analysis, the network's channel-level performance metrics, such as coverage probability and achievable data rate, can be evaluated.

\begin{table*}[ht]
\centering
\caption{History and development of spherical SG (until July 2024).}
\label{tableI-1}
\renewcommand{\arraystretch}{1.1}
%\resizebox{0.6\linewidth}{!}{
\begin{tabular}{|c|c|c|c|c|c|c|}
\hline
Research contents & Phase  &  2020  & 2021 & 2022 & 2023 & 2024  \\ \hline \hline

Basic framework for single-hop coverage & $\mathscr{P}$1 & \cite{200101,200103,200102,200104,200105} & \cite{210101,210102,210103,210104,210105} & \cite{220101,220102,220103,220104,220105,220106,220107} & & \\ \hline

Refined channel modeling and association & $\mathscr{P}$2 & & \cite{212101,212102,212103} & \cite{222101,222102,222103,222104} & \cite{232101,232102,232103,232104,232105,232106,232107,232108} & \cite{242101,242102,242103,242104,242105,242106,242108,242109} \\ \hline

Exploration of complex topologies & $\mathscr{P}$3-1 & & & \cite{222203,222202,222204,222205,222206} & \cite{232201,232202,232203,232205,232206,232207} & \cite{242201,242204,242205,242206,242207,242208,242209,242210,242211} \\ \hline

Integration of new NTPs & $\mathscr{P}$3-2 & & & \cite{222301,222302,222303} & \cite{232301,232302,232303,232304,232305,232306,232307,232308,232309} & \cite{242301,242302,242303,242304,242305} \\ \hline

Researches on advanced topics & $\mathscr{P}$4 & & & \cite{220301,220302} & \cite{230301,230302,230303} & \cite{240301,240302,240303,240304,240305,240306} \\ \hline

Total number of publications & & 5 & 8 & 21 & 26 & 28 \\ \hline
\end{tabular}
%}
\end{table*}

\subsubsection{History and Development}
The concept of spherical SG was first proposed in \cite{gao2019spectrum} and applied to HAP modeling. Although the authors ultimately performed performance analysis by mapping the spherical PPP to a two-dimensional planar PPP, its idea of spherical modeling has provided valuable inspiration for subsequent researchers. In $2020$, two research groups independently developed analytical frameworks for satellite networks based on spherical SG \cite{200103,200105}. From $2020$ to $2024$, a plethora of literature on spherical SG-based NTN analysis was published, with the number of publications increasing significantly each year. This literature can be further divided into four phases (enlisted below). Table~\ref{tableI-1} details the publications for each phase by year.

\begin{figure*}[t]
\centering
\vspace{-0.2cm}
\includegraphics[width = 0.75\textwidth]{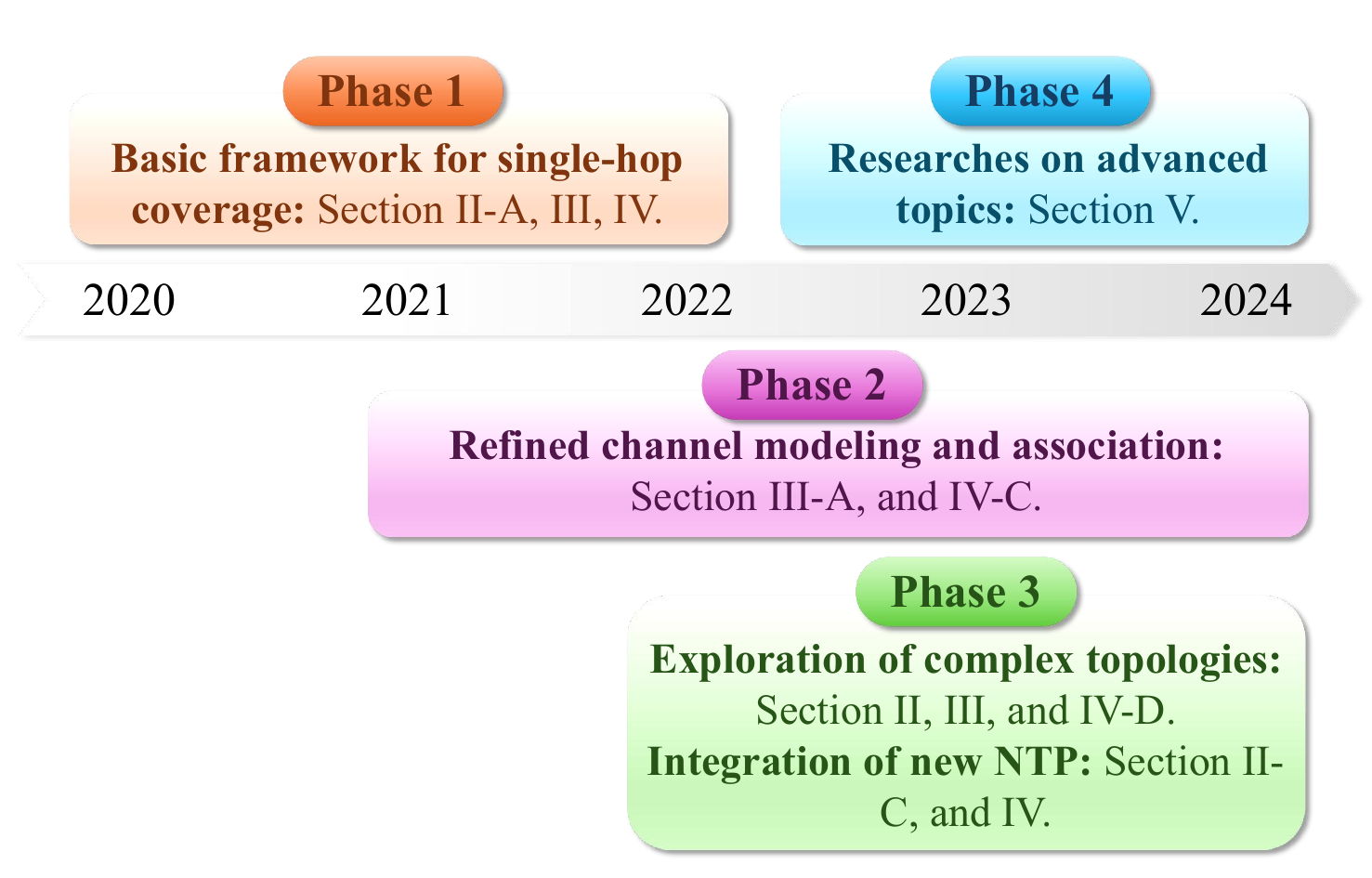}
\caption{Development of spherical SG.}
\vspace{-0.2cm}
\label{figure1-2}
\end{figure*}

\begin{itemize}
    \item Basic framework for single-hop coverage ($\mathscr{P}$1): Scholars have established a fundamental analytical framework for spherical SG concerning the simplest scenario of single-hop coverage for satellite communications.
    \item Refined channel modeling and association ($\mathscr{P}$2): The modeling of the channel is more refined, such as considering the modeling of beam gain and incorporating more strategies for associating with serving NTPs. $\mathscr{P}$2 is a continuation of the study on the simplest scenario of $\mathscr{P}$1 for satellite communications.
    \item Exploration of complex topologies ($\mathscr{P}$3-1); Integration of new NTPs ($\mathscr{P}$3-2): Can be specifically divided into two research branches. Branch $\mathscr{P}$3-1 investigates more complex network topologies, extending from the original single-hop satellite communication to multi-hop scenarios, and expanding from non-orbital satellite constellation models to stochastic and fixed orbital models. Branch $\mathscr{P}$3-2 introduces new equipment such as HAPs, aircraft, ships, and GEO satellites, thereby expanding the research scope of spherical SG. 
    \item Researches on advanced topics ($\mathscr{P}$4): New areas and architectures such as routing strategy design, secure issues, satellite cluster, energy harvesting and satellite-enabled positioning problems are being studied or are yet to be explored. 
\end{itemize}
Finally, Fig.~\ref{figure1-2} shows the development of spherical SG and the section of the article corresponding to each phase.

\subsubsection{Motivation and Advantage}
Now we present the motivation for introducing the spherical SG framework in NTN network performance analysis and its advantages over other research methods.

\par
Unlike most terrestrial communication scenarios, the communication devices in NTN, whether LEO satellites, MEO satellites, or HAPs, are non-stationary relative to the ground. Therefore, it is necessary to consider the dynamic changes in the positions of NTPs during performance evaluation. In the modeling based on the spherical SG framework, the positions of NTPs follow a random distribution, which can effectively meet the needs of a dynamic network topology \cite{220102}.

\par
Additionally, the deployment of NTPs is continuously increasing, such as the expansion of satellite launches, which has been growing by $30\%$ annually since $2012$ \cite{8976900}. As the number of NTPs increases, the impact of co-frequency interference on network performance cannot be ignored. The SG framework is one of the few analytical frameworks with interference analysis capabilities \cite{232302}. 

\par
Finally, compared to other performance evaluation methods such as software-based simulation, the spherical SG analytical framework's most important advantage is its lower computational complexity \cite{232207}. For instance, when conducting satellite constellation simulation and performance evaluation using software tools like systems tool kit (STK), it is necessary to determine the positions of each satellite on different orbits at different times, simulate satellite-to-user communication links, and obtain instantaneous performance. Due to the changing and stochastic nature of satellite positions and link fading, thousands or even more rounds of evaluations are required to accurately obtain the network's average performance \cite{qi2015research}. {\color{black}For the interference performance analysis of LEO constellations, the authors in \cite{232207} state that conducting a round of estimation using the spherical SG-based analytical method only requires 0.458 seconds. In contrast, utilizing state-of-the-art software, even running only 5000 rounds of Monte Carlo simulations would take 1968 seconds, which is 4000 times longer than the analytical method. Furthermore, creating a simulation result figure often involves evaluating performance under dozens or even hundreds of parameter sets, necessitating several days to obtain the results. In fact, many metrics such as coverage probability demand higher precision, rendering 5000 rounds often insufficiently accurate. As a result, performance evaluation methods based on software simulations require substantial computational support.}

%翻译成英文 对于LEO星座的干扰性能分析，作者们申明通过基于球面SG的解析方法进行一轮估计仅需0.458秒。而使用前沿的仿真软件，即使仅运行5000轮蒙特卡洛仿真，也需要1968秒，是解析方法的4000倍。而一张仿真结果图往往需要包含数十甚至上百组参数下的性能评估，因此需要数天才能得到运行结果。事实上，许多指标比如覆盖概率对精度有着更高的要求，5000轮往往不够精确。As a result, 基于软件仿真的性能评估方法requires substantial computational support. 

The computational load for performance evaluation is acceptable for constellations consisting of dozens of satellites. However, for mega-constellations containing thousands of LEO satellites, software-based performance evaluation requires substantial computational support. 

\par
The SG analytical framework considers the randomness of topology and channel fading during distribution and channel modeling, enabling the mapping of network input parameters (such as the number and altitude of satellites) to performance metrics through analytical expressions with low complexity. Since the average performance metrics obtained from analytical expressions are deterministic, only one round of evaluation is required, significantly reducing the computation time for performance evaluation.

\begin{table*}[t]
\centering
{\color{black}
\caption{\color{black}Comparison table with existing surveys.}
\label{tableI-3}
\renewcommand{\arraystretch}{1.1}
\begin{tabular}{|c|c|c|c|c|}
\hline
References & Year & Type & Network category & Research focus \\ \hline \hline
\cite{elsawy2013stochastic} & 2013  & Survey & Terrestrial cellular networks  & Multi-tier and cognitive cellular networks  \\ \hline
\cite{elsawy2016modeling} & 2017  & Tutorial & Terrestrial cellular networks  & Mathmatical modeling and analytical derivation  \\ \hline
\cite{lu2021stochastic} & 2021  & Tutorial  & Terrestrial networks & Spatial-temporal performance analysis \\ \hline
\cite{hmamouche2021new} & 2021  & Survey  & Terrestrial and UAV networks  & New Trends for wireless
networks  \\ \hline
Our work & 2024 & Survey  & Non-terrestrial networks  & Spherical SG models and analytical framework  \\ \hline
\end{tabular}
}
\end{table*}

\subsection{Related Works and Contribution}
This section begins by delineating the distinctions between this survey and other tutorials or surveys on SG frameworks in wireless communication domains. Subsequently, we present the organization and contributions of this survey.

\subsubsection{Related Works}
The establishment of an SG analytical framework for performance analysis in wireless communication networks has been extensively and comprehensively studied. There are several tutorials and surveys in this field, and we list the four most representative ones. In \cite{elsawy2013stochastic}, the authors surveyed the performance analysis of two of the hottest research topics in the field at that time: multi-tier networks and cognitive cellular networks. Subsequently, the paper \cite{elsawy2016modeling} focused on a broader range of applications, namely the modeling and performance analysis of terrestrial cellular networks, and provided a comprehensive tutorial. In \cite{lu2021stochastic}, the authors leveraged the advantages of SG in analyzing spatial randomness to write a tutorial on a novel direction: spatial-temporal performance analysis. One of the most recent tutorials, \cite{hmamouche2021new}, surveyed the latest research in the SG field over the previous decade.

\par
However, in recent years, the field of SG has been undergoing significant transformations, and the emergence of spherical SG has breathed new life into this domain. Currently, none of the existing tutorials or surveys cover the concept of spherical SG, nor do they summarize the applications of SG in NTN. Therefore, a new survey is necessary for updating. This survey focuses more on the unique three-dimensional modeling and topological research of spherical SG, the novel network architectures within NTN, and the future development directions of spherical SG and NTN networks. Instead, the analysis of the wireless channel part generally continues the mathematical techniques used for terrestrial networks, and thus, this aspect will not be elaborated upon in this survey. Consequently, the research presented in this survey significantly differs from current tutorials and surveys and can serve as an important supplement to the SG framework in wireless networks.

\subsubsection{Contribution}
As the first survey in the field of spherical SG, the contributions are as follows:
\begin{itemize}
    \item \textbf{Comprehensive Summary:} In this survey, all spatial distribution models based on spherical SG are introduced, main results at the topological level are derived, and most of the advanced topics are summarized. Advanced topics can be regarded as application scenarios beyond simple uplink and downlink transmission. Advanced topics usually involve various mathematical tools and are not limited to the communication field. 
    \item \textbf{Technical Contribution:} This survey presents original technical contributions to fill the gaps in existing spherical SG research. The definition and generating algorithm of the dual stochastic binomial point process (DSBPP) are provided. A case study on the necessity of spherical modeling is presented, and the error of planar approximation of the spherical model is quantitatively analyzed. For advanced topics, we point out future research directions and provide a case study to fill the research gaps in the existing research.
    \item \textbf{Detailed Classification:} Firstly, we divide the spherical SG-based NTN research into four phases based on their time of appearance, with each phase focusing on different research contents. Secondly, we classify the spherical SG models into three types based on the presence of orbits and whether the modeling of the orbit is stochastic or deterministic. Finally, We classify the links into six categories and provide a detailed explanation of differences in channel fading. The classifications are visually presented through Tables~\ref{tableI-1}, \ref{tableII-1}, and \ref{tableIV-1}, with representative references provided. 
\end{itemize}

\par
\begin{figure}[ht]
\centering
\vspace{-0.2cm}
\includegraphics[width = 0.83 \linewidth]{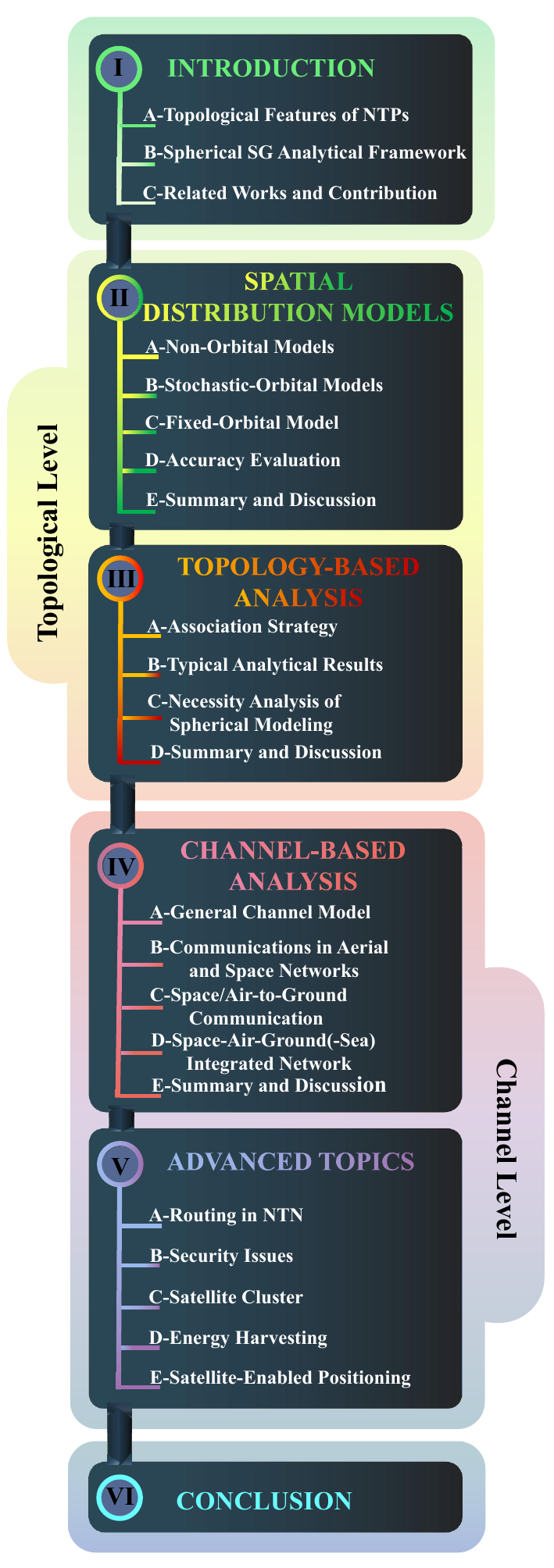}
\caption{Structure of the survey.}
\vspace{-0.2cm}
\label{figure1-3}
\end{figure}

Finally, the abbreviations, along with their corresponding full forms, can be found in Table~\ref{tableI-2}. {\color{black}The structure of the survey is shown in Fig.~\ref{figure1-3}.}

\section{Spatial Distribution Models}\label{section2}
\begin{figure*}[t]
\centering
\vspace{-0.2cm}
\includegraphics[width = 0.85\textwidth]{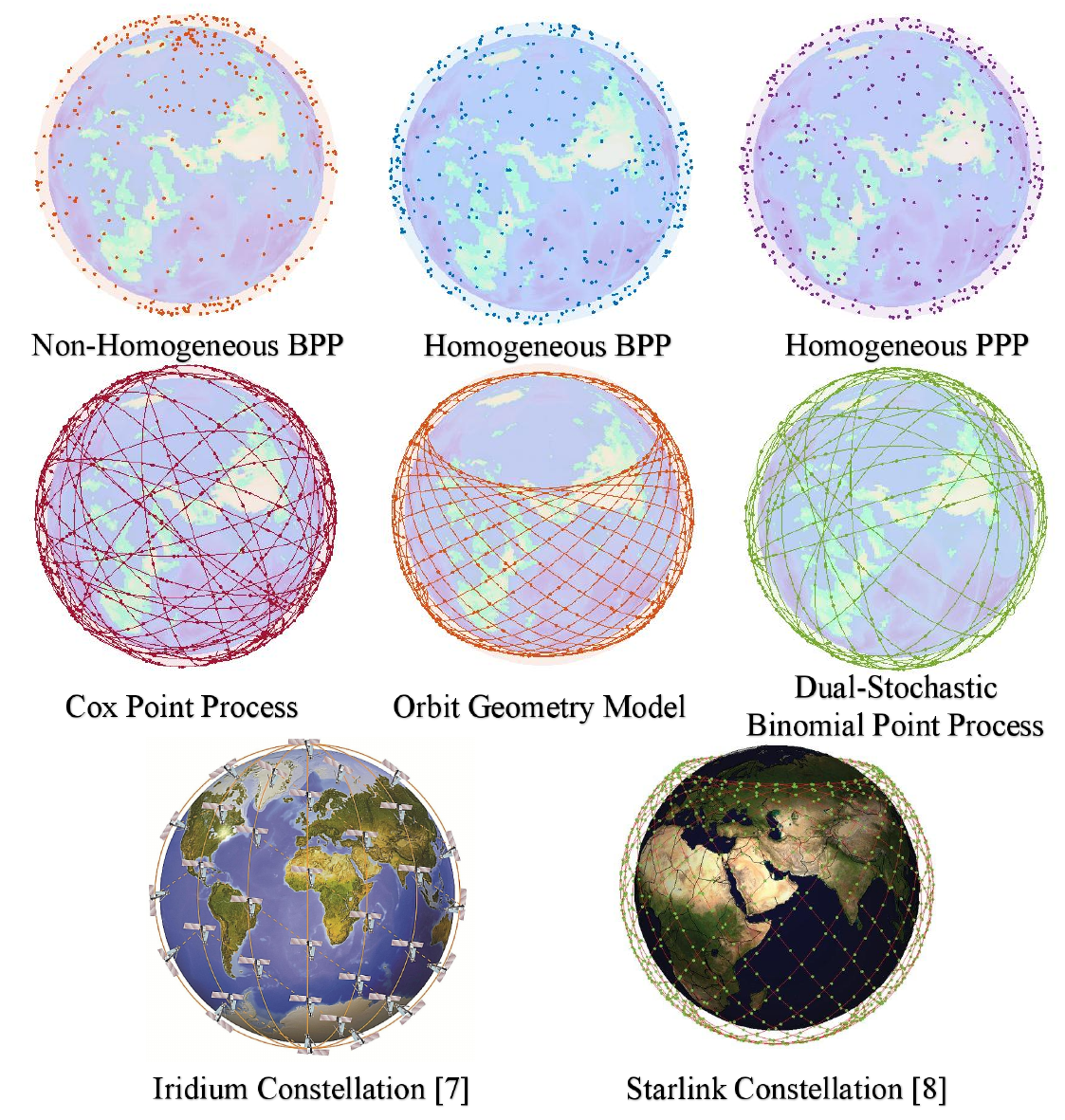}
\caption{Schematic diagrams for constellation configurations.}
\vspace{-0.2cm}
\label{figure2-1}
\end{figure*}

Based on the fact that multiple kinds of NTPs move along fixed trajectories, we introduce three types of spherical SG models: non-orbital models, stochastic-orbital models, and fixed-orbital models. Fig.~\ref{figure2-1} shows examples of these spherical SG-based models. Furthermore, we discuss the accuracy of spherical modeling. Finally, we summarize the differences and relationships between these spherical SG-based models.

\subsection{Non-Orbital Models}\label{sectionII-1}
Among all the models, the non-orbital model offers the highest degree of tractability and is represented in the simplest way. A common feature of these models is that the position of each NTP is independent and identically distributed on a sphere centered on the Earth.

\subsubsection{Homogeneous Poisson Point Process}
The homogeneous Poisson point process (PPP) is one of the most widely used models within the SG framework. In the spherical coordinate system, completely specifying the position of a point requires three coordinates: radius, azimuth angle, and polar angle. In the spherical PPP model, the azimuth angle for each NTP is uniformly distributed, that is, $\varphi_{\mathrm{PPP}} \sim \mathcal{U}[0,2\pi)$. The probability density function (PDF) for the polar angle is given as
\begin{equation}\label{equII-1}
    f_{\theta_\mathrm{PPP}} (\theta) = \frac{\sin\theta}{2}, \ 0 \leq \theta_\mathrm{PPP} \leq \pi,
\end{equation}
which indicates that the density of PPP increases gradually from the poles to the equator \cite{222202}. Note that the PDF of (\ref{equII-1}) can be derived by taking the derivative of (\ref{CDFofsingle}). Although the density is non-uniformly distributed with regard to latitudes, the average number of NTPs per unit sphere surface area remains the same. 

\par
Based on the above analysis, knowing the radius is sufficient to determine the distribution of a platform's position. The number of platforms follows a Poisson process with density $\lambda_{\mathrm{PPP}}$ \cite{210103}. For a network distributed across the entire surface of the sphere, the relationship between intensity $\lambda_{\mathrm{PPP}}$ and number of platforms $N_{\mathrm{PPP}}$ is as follows \cite{210104}:
\begin{equation}
    N_{\mathrm{PPP}} = \lambda_{\mathrm{PPP}} \, 4 \pi R_{\mathrm{PPP}}^2,
\end{equation}
where $R_{\mathrm{PPP}}$ is the radius of the sphere. Planar PPP has been extensively studied in ground-based networks, and many properties remain valid for spherical PPP, such as the Probability generating functional (PGFL) property applicable for interference analysis \cite{232103}.

\subsubsection{Homogeneous Binomial Point Process}
In a homogeneous binomial point process (BPP), the distributions of the azimuth angle and polar angle for each point are consistent with those in a homogeneous PPP. The difference lies in that the number of points in the BPP is constant rather than following a Poisson distribution \cite{220302}. 

\par
Compared to PPP, BPP is more suitable for modeling NTPs on closed spherical surfaces, such as satellite constellations \cite{220102}. Since the number of satellites in the real world is constant, BPP is more accurate than PPP especially when modeling small-scale satellite networks. In the context of SG-based NTP performance analysis, BPP is more commonly used that PPP. In Algorithm~\ref{alg1}, we provide the specific algorithm for generating a BPP on the sphere. 

\begin{algorithm}[!ht] 
    \caption{BPP Generation Algorithm.}
	\label{alg1} 
	\begin{algorithmic} [1]
	\STATE \textbf{Input} Number of points $N_{\mathrm{BPP}}$ and radius $R_{\mathrm{BPP}}$.

    \FOR{$n = 1 : N_{\mathrm{BPP}}$}
    \STATE $u^{(n)} = {\mathrm{rand}}(0,1)$, $v^{(n)} = {\mathrm{rand}}(0,1)$.
    \STATE $\varphi^{(n)} = 2 \pi u^{(n)}$.
    \STATE $\theta^{(n)} = \arccos \left(1- 2 v^{(n)} \right)$.
    \STATE $x^{(n)} = \left(R_{\mathrm{BPP}},\varphi^{(n)},\theta^{(n)} \right)$.
    \ENDFOR
    
	\STATE \textbf{Output}: Positions of points in BPP \\ $\mathcal{X} = \Big\{ x^{(1)},x^{(2)}, \dots, x^{(N_{\mathrm{BPP}})} \Big\}$.
	\end{algorithmic}
\end{algorithm}	
In Algorithm~\ref{alg1} step (3), ${\mathrm{rand}}(0,1)$ generates a number uniformly distributed between $0$ and $1$. To prove step (5), we need to start with the cumulative distribution function (CDF) of the BPP's polar angle,
\begin{equation}
    F_{\theta_{\mathrm{BPP}}} ( \theta) = \frac{1 - \cos\theta}{2},
\end{equation}
which will be derived in (\ref{CDFofsingle}). Let $F_{\theta_{\mathrm{BPP}}} ( \theta)$ follow a uniformly distributed random variale $v \sim \mathcal{U}[0,1]$: 
\begin{equation}
\begin{split}
    F_{\theta_{\mathrm{BPP}}} ( \theta) = \frac{1 - \cos\theta}{2} = v, \\
    \theta = 1 - \arccos\left( 1 - 2v \right),
\end{split}
\end{equation}
and the equation in step (5) follows.

\subsubsection{Non-Homogeneous Point Process}
{\color{black}This part introduces the non-homogeneous point process by comparing it with the homogeneous point process. To provide a clearer distinction between the two, we discuss the differences in detail from three perspectives: mathematical modeling, spatial distribution, and tractability. First, we introduce the differences between the two in terms of mathematical modeling. The non-homogeneous BPP was first proposed by the authors in \cite{200103} and was used to model polar orbit constellations. The only mathematical difference between the non-homogeneous BPP proposed in \cite{200103} and homogeneous BPP is that the distribution of polar angle in non-homogeneous BPP is uniform, i.e., $\theta_{\mathrm{BPP}} \sim \mathcal{U}[0,\pi)$.

\par
Secondly, we use simulation results as examples to demonstrate the differences in distribution between homogeneous and non-homogeneous point processes. The three subfigures at the top of Fig.~\ref{figure2-1} respectively display examples of three non-orbital point processes generated by the aforementioned methods. The non-homogeneous BPP proposed in \cite{200103}, homogeneous BPP, and homogeneous PPP each contain (an average of) $500$ satellites, with the constellation altitude set at $500$~km. By comparison, it is found that the difference between the homogeneous PPP and the homogeneous BPP is minimal. However, compared to the homogeneous BPP, the satellites in the non-homogeneous BPP are noticeably clustered at the Earth's poles, making it a good approximation for the Iridium constellation.

\par
With the construction of the mega LEO satellite network, the shortcomings of the polar orbit constellation configuration are becoming apparent. Polar communication services have relatively low traffic volume, and the dense deployment of satellites at the poles leads to wastage of communication resources. As shown at the bottom of Fig.~\ref{figure2-1}, constellations like Starlink set the orbital inclination to $53$ degrees and build constellation configurations with non-homogeneous distribution along the latitude \cite{korobkov2020traffic}. Driven by this trend, the authors in \cite{210101} proposed a non-homogeneous PPP model. They changed the density of satellite deployment as a function of latitude by altering the polar angle distribution. As a result, non-homogeneous modeling is more realistic and accurate for satellite constellation modeling.

\par
Thirdly, the non-homogeneous point process is far less analytically tractable than the homogeneous one in practice. Performance metrics or intermediate results (such as distance distributions) based on the non-homogeneous point process model will be expressed as functions of latitude, significantly increasing the complexity of expressions \cite{220101}. In contrast, since the homogeneous point process has the same distribution for users at any location, the expression for the distance distribution is significantly simpler than that of the non-homogeneous point process. Faced with topologically complex situations such as satellite routing problems, the non-homogeneous model may render routing performance almost intractable. 
}

\subsection{Stochastic-Orbital Models}
Considering that non-orbital models ignore the fact that satellites and airplanes move along predetermined trajectories, modeling such NTPs as a BPP or a PPP might be considered less accurate for continuous-time observations \cite{242205}. Therefore, researchers have proposed the spherical stochastic-orbital models under the SG framework.

\subsubsection{Cox Point Process}
The stochastic-orbital model indicates that the trajectory follows a specific stochastic distribution relative to the position of a typical user. So far, the CPP is the only proposed stochastic-orbital model. To generate the CPP, the authors in \cite{222205} first consider a two dimensional PPP of density $\frac{\lambda_{\mathrm{orb}} \sin(\theta_{\mathrm{inc}})}{2\pi}$ on rectangle set $[0,\pi) \times [0,\pi)$ as the orbit process, where $\lambda_{\mathrm{orb}}$ is the average number of orbits in the constellation. The coordinate of each point on the rectangle set is denoted as $(\theta_{\mathrm{lon}}, \theta_{\mathrm{inc}})$. Specifically, $\theta_{\mathrm{lon}}$ is the longitude and $\theta_{\mathrm{inc}}$ is the inclination angle. Then, satellites on each orbit follow a mutually independent PPP with density $\frac{\lambda_{\mathrm{sat}}}{2\pi}$, where $\lambda_{\mathrm{sat}}$ is the average number of satellites on a single orbit \cite{242209}. Finally, the positions of satellites are mapped to the sphere. We refer to the set of satellite positions on all orbits as the satellite point process. Since the satellite point process is conditionally defined on the orbit process, it is known as a CPP.

\par
Performance analysis based on the CPP is more complex than that based on the non-orbital models. Therefore, in order to minimize the analytical complexity, the orbit point process and satellite point process of the CPP are carefully designed as explained above to meet homogeneity \cite{242206}. Although the introduction of the CPP model has brought groundbreaking advancements to the spherical SG field, CPP may not necessarily be more accurate compared to the non-orbital model. This is because both the number of orbits and the number of satellites on each orbit are assumed to follow a Poisson distribution rather than a deterministic value, and the randomness of the Poisson distribution cannot be ignored. For example, as a mega-constellation, Starlink has only $22$ satellites per orbit \cite{liang2021phasing}. In CPP, when the number of satellites follows a Poisson distribution with a mean of $22$, the randomness in the number of satellites may lead to significant deviations between the estimated results and actual performance. Finally, due to the differences between the Poisson distribution and the actual constellation in modeling, it is not easy to intuitively understand the modeling process of CPP.

\subsubsection{Dual Stochastic Binomial Point Process}
\begin{figure}[ht]
\centering
\vspace{-0.2cm}
\includegraphics[width = \linewidth]{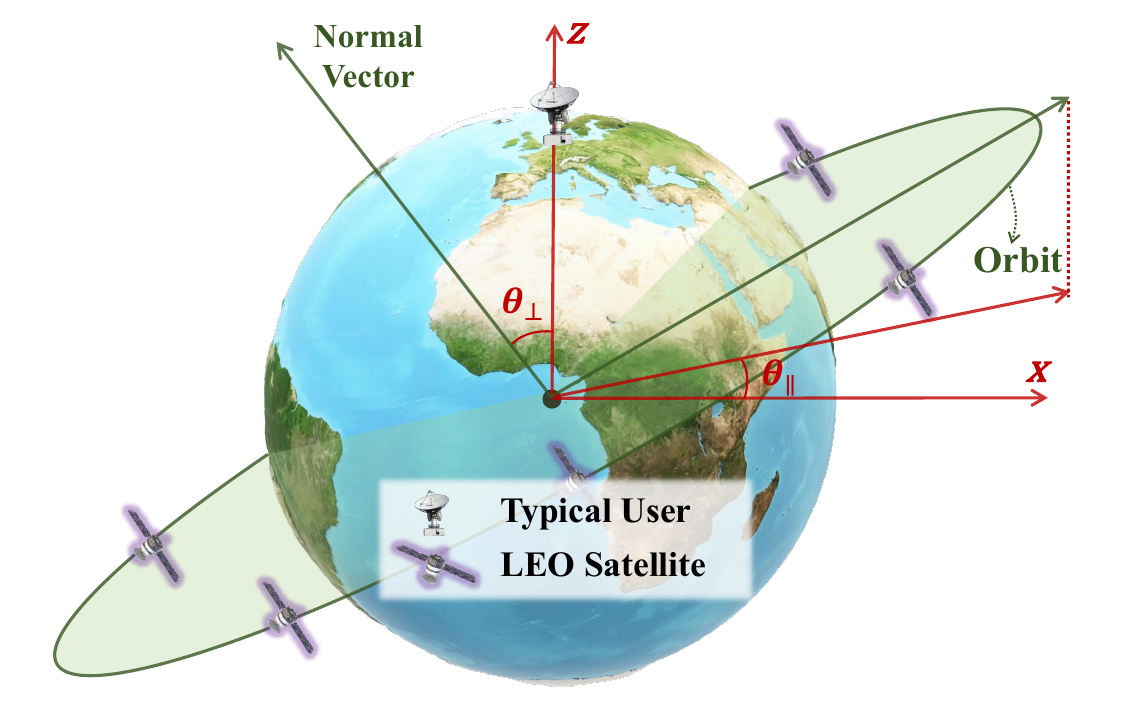}
\caption{Diagram for the orbital-based model.}
\vspace{-0.2cm}
\label{figure2-2}
\end{figure}

Considering the aforementioned shortcomings of the CPP, another potential approach is to consider a DSBPP. Like CPP, DSBPP also belongs to the category of stochastic-orbital models. The main difference between DSBPP and CPP is that the number of orbits and the number of satellites in each orbit are fixed values in DSBPP. Specifically, orbits in DSBPP are also determined by two parameters: inclination angle $\theta_{\perp}$ and azimuth angle $\theta_{\parallel}$. As shown in Fig.~\ref{figure2-2}, the inclination angle $\theta_{\perp}$ is defined as the angle between the positive direction of the $z$-axis and the normal vector of the orbit plane. The PDF of $\theta_{\perp}$ is given as,
\begin{equation}
    f_{\theta_{\perp}} (\theta) = \frac{\sin\theta}{2}, \ 0 \leq \theta_{\perp} \leq \pi.
\end{equation}

\par
Fig.~\ref{figure2-2} also provides an example of the azimuth angle $\theta_{\parallel}$. 
We first select an arbitrary point on the positive half of the $z$-axis, excluding the origin, and find the point on the orbital plane that is closest to this selected point, denoted as $\epsilon$. Project the line between the origin and $\epsilon$ on the $xy$ plane, the angle between the projection and the positive direction of the $x$-axis is the azimuth angle $\theta_{\parallel}$. We assume the azimuth angle is uniformly distributed, that is $\theta_{\parallel} \sim \mathcal{U}[0,2\pi)$.

Compared to CPP, the modeling process of DSBPP is easier to understand. Referring to the generation process of CPP \cite{222205}, it is not difficult to prove that DSBPP is also homogeneous. In Fig.~\ref{figure2-1}, we generated two constellations, each with (an average of) $25$ orbits and (an average of) $20$ satellites per orbit, at an altitude of $500$~km: one for CPP and one for DSBPP. As shown in the figure, the differences between CPP and DSBPP are not significant based on the generated cases. Finally, in Algorithm~\ref{alg2}, we provide pseudocode for the generation of DSBPP. Unlike Algorithm~\ref{alg1}, which is applicable to any programming language, in Algorithm~\ref{alg2} we use MATLAB syntax. MATLAB is well-suited for matrix operations and vectorized computations, and it also offers a rich set of mathematical functions.

\begin{algorithm}[!ht] 
    \caption{DSBPP Generation Algorithm.}
	\label{alg2} 
	\begin{algorithmic} [1]
	\STATE \textbf{Input}:  \textit{NumOrbit} (Number of orbits), \textit{NumSat} (number of satellites on a single orbit), and \textit{OrbitRadius} (radius of orbits). 

    \STATE \textbf{Initiate}: ${\mathrm{\textit{SatPos}}} = {\mathrm{zeros}} (3, {\mathrm{\textit{NumOrbit}}} * {\mathrm{\textit{NumSat}}})$, and ${\mathrm{\textit{Index}}} = 1$.

    \STATE ${\mathrm{\textit{Incs}}}={\mathrm{acos}}(1-2*{\mathrm{rand}}(1,\textit{NumOrbit}))/\pi*180$.

    \STATE ${\mathrm{\textit{Azims}}}={\mathrm{rand}}(\textit{NumOrbit},1)*360$.

    \FOR{$n = 1 : {\mathrm{\textit{NumOrbit}}}$}
    \STATE ${\mathrm{\textit{SatAngles}}} = {\mathrm{rand}}({\mathrm{\textit{NumSat}}}, 1) * 360$.
    \FOR{$m = 1 : {\mathrm{\textit{NumSat}}}$}

    \STATE $\textit{x} = {\mathrm{\textit{OrbitRadius}}} * {\mathrm{cos}}( {\mathrm{\textit{SatAngles}}}(m) )$.

    \STATE $\textit{y} = {\mathrm{\textit{OrbitRadius}}} * {\mathrm{sin}}( {\mathrm{\textit{SatAngles}}}(m) )$.

    \STATE $\textit{z} = 0$.

    \STATE $\textit{Rx} = [1, 0, 0; 0, {\mathrm{cosd}}( {\mathrm{\textit{Incs}}} (n)), - {\mathrm{sind}}( {\mathrm{\textit{Incs}}} (n));0, \ \ $  ${\mathrm{sind}}({\mathrm{\textit{Incs}}} (n)), {\mathrm{cosd}}({\mathrm{\textit{Incs}}} (n))]$.
    
    \STATE $\textit{Rz} = [{\mathrm{cosd}}({\mathrm{\textit{Azims}}}(n)), -{\mathrm{sind}}({\mathrm{\textit{Azims}}}(n)), 0; \ \ \ \ \ \ $ ${\mathrm{sind}}({\mathrm{\textit{Azims}}}(n)),  {\mathrm{cosd}}({\mathrm{\textit{Azims}}}(n)), 0; 0, 0, 1]$.

    \STATE $\mathrm{\textit{SatLoc}} = \textit{Rz} * (\textit{Rx} * [\textit{x}; \textit{y}; \textit{z}]) $.

    \STATE ${\mathrm{\textit{SatPos}}} (:, {\mathrm{\textit{Index}}}) = {\mathrm{\textit{SatLoc}}}$.
    \STATE ${\mathrm{\textit{Index}}} = {\mathrm{\textit{Index}}} + 1$.
    \ENDFOR
    \ENDFOR
    
	\STATE \textbf{Output}: \textit{SatPos} (Positions of satellites).
	\end{algorithmic}
\end{algorithm}
In Algorithm~\ref{alg2}, 
all variables are in italics. Note that \textit{SatPos}, \textit{Inclinations}, \textit{Azimuths}, \textit{SatAngles} and \textit{SatLoc} are vectors, \textit{Rx} and \textit{Rz} are matrices. Steps (3), (4), and (6) in Algorithm~\ref{alg2} randomly generate the inclination angle for each orbit, azimuth angle for each orbit, and position for each satellite in the $n^{th}$ orbit. In step (8), ${\mathrm{\textit{SatAngles}}}(m)$ is the $m^{th}$ element in vector \textit{SatAngles}. Steps (11)-(13) map a set of random angles to a satellite's coordinates in the Cartesian coordinate system by introducing rotation matrices.

\subsection{Fixed-Orbital Model}
Unlike the stochastic-orbital model, where the orbital process follows a specific random distribution, the fixed-orbital model has deterministic orbit positions. The spherical fixed-orbital models can be currently classified into the Poisson line process (PLP) and the orbit geometry model (OGM). PLP can be regarded as modeling the NTP on a single orbit in CPP, and the inclination angle of this orbit is given \cite{242102}. The OGM is a combination of multiple PLPs with the same center and radius, and each orbit plane of the PLP has a given inclination angle. 

\par
Thus, it is evident that the inclinations in the fixed-orbital models are known and do not follow a random distribution. Therefore, network performance metrics based on these models are functions of the inclination angles, and all current fixed-orbital models do not satisfy homogeneity \cite{222202}. As a result, expressions derived from such models are complex, trading off analytical simplicity for modeling accuracy. In Fig.~\ref{figure2-1}, we modeled a constellation at an altitude of $500$~km with $25$ orbits and $20$ satellites per orbit using the OGM. Each orbit in the OGM has an inclination angle of $53$~degrees, and the azimuth angles of the orbits are distributed with equal space. Compared to the stochastic-orbital models (CPP and DSBPP), OGM more closely resembles the Starlink constellation \cite{korobkov2020traffic}.

\subsection{Accuracy Evaluation}
From the previous descriptions, it can be seen that several SG-based models trade varying degrees of modeling accuracy for analytical tractability. Therefore, the accuracy of spherical modeling has been a topic of debate.

\par
So far, numerous studies have demonstrated that the spherical SG analytical framework can achieve performance evaluation results close to those obtained from actual NTNs. In \cite{200101}, the authors compared the satellite network interference estimation results under the spherical PPP model with those obtained from STK software simulations, proving the accuracy of the spherical PPP in interference analysis. Furthermore, in \cite{200104}, the authors showed that the coverage probability estimated by the spherical BPP model is almost identical to the results obtained from the deterministic Walker constellation configuration, with only limited deviations in data rate estimates. {\color{black} The author in \cite{212101} and \cite{212102} also provided a comparison with the Walker constellation configuration, and proved that the pass duration estimated by the spherical SG framework is accurate.} Additionally, in \cite{232308}, the authors used the PLP to model the GEO constellation and analyzed the distance distribution from a typical ground user to the nearest GEO satellite. The results indicated that at any latitude, the distance distribution under the PLP model coincides with the actual distance distribution in the GEO constellation. Finally, the CPP model was also proven to provide consistent coverage probability estimates with the Starlink constellation \cite{242206}.

\par
Although multiple models have been qualitatively proven to achieve performance evaluation results close to those of actual constellations, they all assume that the constellation comprises at least several hundred satellites. For smaller-scale constellations, the accuracy of applying spherical SG-based models needs to be quantitatively assessed. Therefore, we propose using the Wasserstein distance as a quantitative measure of the similarity between two point processes \cite{222202}. The physical meaning of the squared Wasserstein distance is the minimum energy required to move one point process to another. In addition to demonstrating the trend of decreasing Wasserstein distance with an increasing number of satellites, the results in \cite{222202} also indicate that the lower the satellite altitude and the smaller the orbital inclination angle, the higher the accuracy of the BPP modeling. {\color{black}Finally, Table~\ref{tableII-2} summarizes the literature on the accuracy analysis of spherical SG Models.}

\begin{table}[ht]
\centering
{\color{black}
\caption{\color{black}Summary of literature on the accuracy analysis of spherical SG Models.}
\label{tableII-2}
\resizebox{\linewidth}{!}{
\renewcommand{\arraystretch}{1.1}
\begin{tabular}{|c|c|c|c|}
\hline
Ref. & Model & NTP & Metric \\ \hline \hline
\cite{200101} & PPP & LEO satellite & Interference \\ \hline
\cite{200104} & BPP  & LEO satellite  & Coverage probability, data rate \\ \hline
\cite{212101} & PPP  & LEO satellite  & Pass duration \\ \hline
\cite{212102} & PPP  & LEO satellite  & Pass duration \\ \hline
\cite{222202} & BPP, OGM & LEO satellite & Wasserstein distance \\ \hline
\cite{242206} & CPP & LEO satellite & Coverage probability  \\ \hline
\cite{232308} & PLP  & GEO satellite  & Distance distribution  \\ \hline
\end{tabular}
}
}
\end{table}

\begin{table*}[t]
\centering
\caption{Representative references for spherical SG-based models.}
\label{tableII-1}
\renewcommand{\arraystretch}{1.1}
\begin{tabular}{|c|c|c|c|c|}
\hline
References & Devices & Model &  Classification & Homogeneous \\ \hline \hline
\cite{200102,200104} & LEO satellite  & BPP  & Non-orbital model  & Non-homogeneous  \\ \hline
\cite{210101,220101} & LEO satellite  & PPP  & Non-orbital model  & Non-homogeneous  \\ \hline
\cite{222101,220302} & LEO satellite  & BPP  & Non-orbital model  & Homogeneous  \\ \hline
\cite{210103,210105} & LEO satellite  & PPP  & Non-orbital model  & Homogeneous  \\ \hline
\cite{232203} & LEO satellite  & PCP  & Non-orbital model  & Non-homogeneous  \\ \hline
\cite{242209,242205} & LEO satellite  & CPP  & Stochastic-orbital model  & Homogeneous  \\ \hline
\cite{242102} & LEO satellite  & PLP   & Fixed-orbital model  & Non-homogeneous  \\ \hline
\cite{222202,222206} & LEO satellite  & OGM   & Fixed-orbital model  & Non-homogeneous  \\ \hline
\cite{232306,242302} & HAP  & BPP  & Non-orbital model  & Homogeneous \\ \hline
\cite{232304} & HAP  & PPP  & Non-orbital model  & Homogeneous  \\ \hline
\cite{232308} & GEO satellite  & PLP  & Fixed-orbital model  & Non-homogeneous \\ \hline
\cite{232301,240302} & Aircraft  & PLP  & Fixed-orbital model  & Non-homogeneous  \\ \hline
\end{tabular}
\end{table*}

\subsection{Summary and Discussion}
Here, we summarize several important conclusions about the point process discussed in this section.
\begin{itemize}
    \item Spherical SG-based point processes can be classified into the non-orbital model, stochastic-orbital model, and fixed-orbital model. The accuracy of these three types of models increases successively, while the difficulties of deriving analytical expressions based on these models also increase accordingly. When considering the impact of continuous motion of NTPs on performance metrics, the non-orbital model is no longer applicable. However, in scenarios involving complex topology analysis, such as multi-hop communication, the non-orbital model remains the only choice in current research due to its strong tractability.
    \item When the network scale is small, the BPP is more suitable for modeling closed surfaces compared to the PPP. However, as the number of NTPs increases, the difference between the two diminishes. Therefore, some studies modeled satellites using BPP but approximated it as PPP during analysis, and some characteristics of PPP, such as void probability can be applied \cite{222104}. The above conclusions can also be extended to the relationship between CPP and DSBPP. 
    \item The performance metrics under the non-homogeneous models are expressed as functions of the orbit inclination angle or latitude, resulting in more complex analytical expressions. Although non-homogeneous models are more accurate compared to homogeneous ones, the latter are more widely applied due to their strong analytical tractability. Both non-homogeneous and homogeneous models are encompassed in non-orbital models. Currently, all stochastic-orbital models are homogeneous, whereas fixed orbital models are non-homogeneous by nature.
    \item The modeling accuracy of SG-based models increases with the number of NTPs. This explains why satellites and HAPs attracted attention early on, but the spherical SG analytical framework has recently been applied with the establishment of mega-constellations.
\end{itemize}

\section{Topology-Based Analysis}\label{section3}
Compared to other studies in the SG field, the spherical SG framework has unique characteristics in topological analysis. In this section, we discuss association strategies, contact distance, and contact angle distribution, as well as availability probability. The latter two are specific research topics within the spherical SG analytical framework. Finally, we introduce a case study to compare the advantages and disadvantages of spherical models versus planar models and discuss the necessity of spherical modeling.

\subsection{Association Strategy}\label{section3-1}
Association refers to the process by which users select the NTPs that provide them with services, while the received power from unassociated NTPs is considered interference. Therefore, the choice of association strategy directly determines the received power and interference power, significantly impacting communication performance.

\par
First, we introduce conditions in which the NTP can be associated. Firstly, due to the high altitude deployment of NTP devices, it is necessary to consider the curvature of the Earth. Therefore, only NTPs above the horizon at the user's location can be associated. Secondly, considering the long communication distance, NTPs usually need to use beamforming technology to directionally transmit signals. Users or NTPs outside the main lobe of the beam or the half-power beamwidth cannot establish a link \cite{242302}. Third, some studies have artificially defined a maximum reliable distance for association, which is determined by factors such as transmission power and channel parameters \cite{220102}.

\par
Next, we discuss association strategies in different scenarios. The simplest scenario is single-hop communication in a single-tier network (all satellites have the same transmit power, altitude, antenna gain, and so on). Structurally, networks can be classified into cellular networks and cell-free networks. In a cellular network, the maximum average received power association strategy is adopted, meaning that the user selects the closest NTP in the single-tier network. In a cell-free network, the user randomly selects an NTP to access, known as the random association strategy \cite{232309}. Compared to the closest NTP association strategy, the random association strategy requires NTPs to have wider beam main lobes and larger service ranges. The communication distance from the associated NTP to the user is also longer, resulting in greater attenuation. In a multi-tier heterogeneous network, a more distant or higher-altitude NTP may provide greater received power to the user when it has higher transmission power or antenna gain. Therefore, the maximum average received power association strategy and the nearest NTP association strategy may result in different choices of the associated NTP \cite{242103} in a multi-tier network.

\par
Association strategies in dual-hop relay communication are far more complex than those in single-hop communication. Authors in \cite{242211} proved that associating with the nearest LEO satellite as a relay between the user and the GEO satellite is the optimal association strategy when the transmission power of ground users is limited. The aforementioned transmitter end nearest NTP association strategy is also adopted in \cite{222204}. In contrast, authors in \cite{242208} applied the receiver end association strategy in terrestrial-satellite-terrestrial relay communication. This is primarily due to considerations of energy efficiency, as satellite onboard energy is more costly compared to the ground.

\subsection{Typical Analytical Results}\label{sectionIII-B}
In this subsection, we introduce and derive some typical analytical results involved in most spherical SG studies. The relationships among these typical analytical results are also revealed.

\subsubsection{Central Angle and Zenith angle}
\begin{figure}[ht]
\centering
\vspace{-0.2cm}
\includegraphics[width = \linewidth]{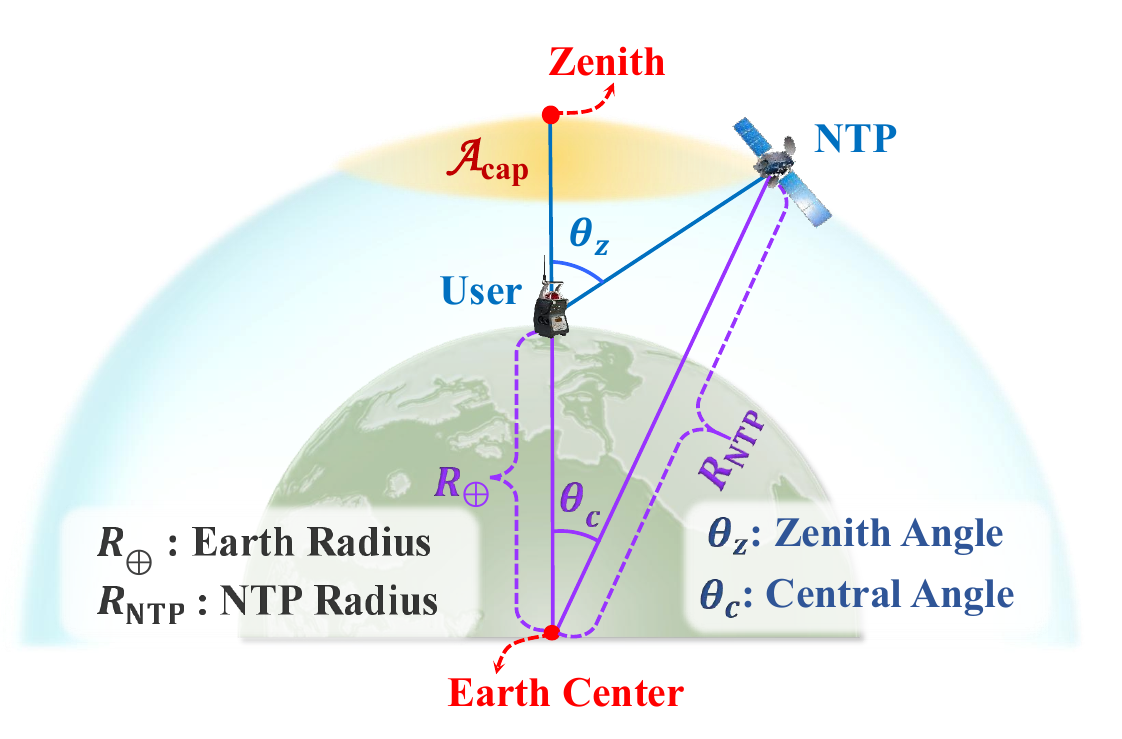}
\caption{A diagram of the central angle and the zenith angle.}
\vspace{-0.2cm}
\label{figure3-1}
\end{figure}

As shown in Fig.~\ref{figure3-1}, we use the case of NTP-to-ground communication to explain the concepts of central angle and zenith angle. The central angle is the angle formed with the Earth center as the vertex, and with the lines connecting the Earth center to the user and the Earth center to the NTP as the sides \cite{220302}. The central angle is widely used because expressing distance through the central angle is more convenient in the spherical coordinate system. Denoting the Euclidean distance from the user to the NTP as $d_{\mathrm{u-n}}$, the relationship between the central angle $\theta_c$ and $d_{\mathrm{u-n}}$ can be expressed by the cosine rule:
\begin{equation}\label{dis2angle}
    d_{\mathrm{u-n}} = \sqrt{R_{\mathrm{NTP}}^2 + R_{\oplus}^2 - 2R_{\mathrm{NTP}}R_{\oplus} \cos\theta_c},
\end{equation}
where $R_{\oplus}$ is the Earth radius, and $R_{\mathrm{NTP}}$ is the radius of the sphere where NTPs are located.

\par
The zenith angle is the angle formed with the user's location as the vertex and with the lines connecting the user to the zenith and the user to the NTP as the sides. An example of the zenith angle is given in Fig.~\ref{figure3-1}.
{\color{black}The relationship between the central angle $\theta_c$ and the zenith angle $\theta_z$ obtained by the Sine rule \cite{210104} 
\begin{equation}\label{centraltozenith}
\begin{split}
    & \frac{R_{\mathrm{NTP}}}{\sin\theta_z} = \frac{1}{\sin\theta_c} \frac{R_{\mathrm{NTP}}\cos\theta_c - R_{\oplus}}{\cos\theta_z}, \\
    \theta_z = & \cot^{-1} \left( 
    \cot\theta_c - \frac{R_{\oplus}}{R_{\mathrm{NTP}}} \sqrt{1 + \cot^2\theta_c} \right).
\end{split}
\end{equation}
}
From the above analysis, we can find that the zenith angle can also establish a one-to-one mapping with distance $d_{\mathrm{u-n}}$. This mapping is indirectly established through the central angle, thus it is not straightforward. Although less convenient for distance expression, the zenith angle can better describe the user's perspective, for example, in representing the beam direction of user-to-NTP uplink transmission. 

\par
Finally, it is worth to notice that 
the rotation of the coordinate system is an important technique when applying the central angle and zenith angle based on a homogeneous point process for analysis. According to Slivnyak's theorem \cite{feller1991introduction}, the rotation of the coordinate system does not affect the distribution of the homogeneous point process. Therefore, we generally rotate the coordinate system to fix the typical user at the north pole, as shown in Fig.~\ref{figure3-1}. In this case, the central angle between the NTP and the user becomes the polar angle of the NTP. As a result, the coordinate of the NTP can be easily calculated using the central angle, which will facilitate derivations.

\subsubsection{Contact Angle and Contact Distance}
In existing research, selecting the nearest NTP to provide services is the most widely adopted association strategy. Therefore, researchers define the central angle between the user and the nearest NTP as the contact angle \cite{222204}, used to measure the distance from the associated NTP to the user. 

\par
Then, the contact angle distribution based on homogeneous BPP is derived. We start with deriving the probability that an NTP is located within the orange spherical cap at the top of Fig.~\ref{figure3-1}:
\begin{equation} \label{CDFofsingle}
    P^{(1)} (\theta_c) = \frac{\mathcal{A}_{\mathrm{cap}}}{4\pi R_{\mathrm{NTP}}^2} = \frac{ 2\pi R_{\mathrm{NTP}}^2 (1-\cos\theta_c)}{4\pi R_{\mathrm{NTP}}^2} = \frac{ 1 - \cos\theta_c}{2},
\end{equation}
where $\mathcal{A}_{\mathrm{cap}}$ is the area of the spherical cap.  Considering that the probability of the NTP in Fig.~\ref{figure3-1} being associated is equal to the probability that there are no NTPs in the spherical cap, the CDF of the contact angle distribution is
\begin{equation}
\begin{split}
    & F_{\theta_0} (\theta) = \mathbbm{P}\left[ \theta \leq \theta_0 \right] = 1 - \left( 1 - P^{(1)} (\theta) \right)^{N_{\mathrm{NTP}}} \\
    & = 1 - \left( 1 - \frac{ 1 - \cos\theta}{2} \right)^{N_{\mathrm{NTP}}} = 1 - \left( \frac{ 1 + \cos\theta}{2} \right)^{N_{\mathrm{NTP}}},
\end{split}
\end{equation}
where $\theta_0$ denotes the contact angle, and $N_{\mathrm{NTP}}$ is the number of NTPs. By differentiating the above CDF, the expression for the PDF of the contact angle distribution can be obtained,
\begin{equation}
    f_{\theta_0} (\theta) = \frac{N_{\mathrm{NTP}}}{2} \sin\theta \left( \frac{ 1 + \cos\theta}{2} \right)^{N_{\mathrm{NTP}}-1}.
\end{equation}

\par
Furthermore, we can also define the contact distance, which is the Euclidean distance between the typical user and the nearest NTP \cite{200103}. {\color{black}Based on the relationship between the Euclidean distance and central angle given in (\ref{dis2angle}), the CDF of contact distance is given as,
\begin{equation}
\begin{split}
    & F_{d_0} (d) = 1 - \left( \frac{1}{2} + \frac{\cos\theta_0(d)}{2}  \right)^{N_{\mathrm{NTP}}} \\
    & = 1 - \left( \frac{1}{2} + \frac{R_{\mathrm{NTP}}^2 + R_{\oplus}^2 - d^2}{4 R_{\mathrm{NTP}} R_{\oplus} }  \right)^{N_{\mathrm{NTP}}}.
\end{split}
\end{equation}
where the relationship between $d$ and $\cos\theta_0(d)$ is established by the Cosine rule.} Compared to the contact angle distribution, the contact distance distribution appears more complex in its expression, but it directly reveals the relationship between contact distance and  
$R_{\mathrm{NTP}}$.

\subsubsection{Availability Probability}
The availability probability is the probability that at least one NTP meets the conditions to be associated with the typical user. The availability probability under the homogeneous BPP model can be derived as,
\begin{equation}
\begin{split}
    & P_1^A = 1 - \left( 1 - P^{(1)} (\theta_c^{\max}) \right)^{N_{\mathrm{NTP}}} \\
    & = 1 - \left( \frac{ 1 + \cos\theta_c^{\max}}{2} \right)^{N_{\mathrm{NTP}}},
\end{split}
\end{equation}
where $\theta_c^{\max}$ is the maximum central angle between the associable NTP and the user. Comparing the analytical expression of the availability probability with the CDF of the contact angle distribution reveals a close relationship between them. 

\par
Next, we extend the scenario of having one associable NTP to multiple. The $K-$availability probability is defined as the probability that at least $K$ NTPs meet the conditions to be associated. {\color{black} Based on the expression of the availability probability, the $K-$availability probability is expressed as
\begin{equation}
\begin{split}
    P_K^A & = 1 - \sum_{k=0}^{K-1} \mathbbm{P} \left[ \mathcal{N} \left( \mathcal{S}(\theta_c^{\max}) \right) = k \right] \\
    & = 1 - \sum_{k=0}^{K-1} \binom{N_{\mathrm{NTP}}}{k} \left( \frac{ 1 - \cos\theta_c^{\max} }{2} \right)^k \\
    & \times \left( 1 - \frac{ 1 - \cos\theta_c^{\max} }{2} \right)^{N_{\mathrm{NTP}}-k},
\end{split} 
\end{equation}
where $\mathcal{S}(\theta)$ represents the spherical cap with a central angle of $2\theta$ and $\mathcal{N} \left( \mathcal{S}(\theta) \right)$ counts the number of LEO satellites in the sperical cap $\mathcal{S}(\theta)$. The second step in the above equation follows the CDF of the binomial distribution.
}

\par
Although the concept of $K-$availability has not yet been mentioned in the existing spherical SG literature, it is common for multiple NTPs to serve one user. For instance, satellite positioning requires the simultaneous participation of at least three satellites.

\subsection{Necessity Analysis of Spherical Modeling}
Now that some mathematical foundations are established, we can better discuss an important issue. Although the spherical SG framework is gaining increasing attention, there is ongoing debate about whether it is necessary to further study the spherical SG model for NTN communications, given that the planar SG model has already been well-developed in terrestrial networks. Here are comparisons between the planar and spherical models. In terms of topological analysis, spherical models are far more complex than planar models. As a result, authors first applied spherical modeling for HAPs in \cite{gao2019spectrum}, and then simplified the derivation by a planar approximation. In contrast, planar models are less accurate compared to the spherical ones since they ignore the curvature of the Earth. The deployment area of the spherical model NTN is consistent with the real-world NTN, with only the position of each NTP approximated. In contrast, the planar model approximates both the deployment area and the position of each NTP and thus can be considered an approximation of an approximation. In summary, spherical modeling is necessary when the difference between spherical and planar models is significant enough. 

\subsubsection{Mapping Rule}
In this survey, we design a case study to investigate the necessity of spherical modeling by comparing the difference of modeling the same NTN using spherical and planar models. First of all, we generate a spherical homogeneous BPP containing $100$ NTPs ($N_{\mathrm{NTP}} = 100$) within a spherical cap with radius $R_{\mathrm{NTP}}$ as the first approximation of the NTP distribution. We denote $2\theta_c$ as the central angle of the spherical cap in Fig.~\ref{figure3-1}, which is also the $x$-axis in Fig.~\ref{figure3-2}. Next, we map the NTPs in the spherical SG model onto a planar homogeneous BPP. The planar BPP is located on a circular plane with radius $R_{\mathrm{NTP}} \sin\theta_c$. The line connecting the Earth center and the zenith passes through the center of the circular plane and is perpendicular to the plane. 

\par
After explaining the shape of the planar, the specific mapping rules are as follows. The number of points in the planar BPP is also $N_{\mathrm{NTP}}$, and the position of each NTP in the planar BPP has a one-to-one correspondence to the spherical one. The position of each NTP in the planar BPP is represented by cylindrical coordinates. Both BPPs share the same set of azimuth angle values, with each NTP's azimuth angle independently following a uniform distribution between $0$ and $2\pi$. 
Subsequently, we generate a set of random numbers
\begin{equation}
    \left\{ u^{(1)}, u^{(2)}, \dots, u^{\left( N_{\mathrm{NTP}} \right)} \right\},
\end{equation}
where $u^{(n)}$ follows a uniform distribution between $0$ and $1$, for $\forall \, 1 \leq n \leq N_{\mathrm{NTP}}$. The polar angle of the $n^{th}$ point in the spherical BPP is given as,
\begin{equation}
    \theta_{\mathrm{spherical}}^{(n)} = \arccos\left(1-u^{(n)} \left(1-\cos\theta_c\right) \right).
\end{equation}
The radial distance of the $n^{th}$ point in the planar BPP is mapped as,
\begin{equation}
    \rho_{\mathrm{planar}}^{(n)} = \sqrt{u^{(n)}} R_{\mathrm{NTP}} \sin\theta_c.
\end{equation}
The above mapping process of the polar angle ensures that both BPPs are homogeneous. Finally, the altitude of the plane $h_{\mathrm{planar}}$, which is defined as the distance from the plane center to the Earth center, satisfies $R_{\mathrm{NTP}} \cos\theta_c < h_{\mathrm{planar}} < R_{\mathrm{NTP}}$.

\subsubsection{Results Analysis}
\begin{figure}[ht]
\centering
\vspace{-0.2cm}
\includegraphics[width = 0.8\linewidth]{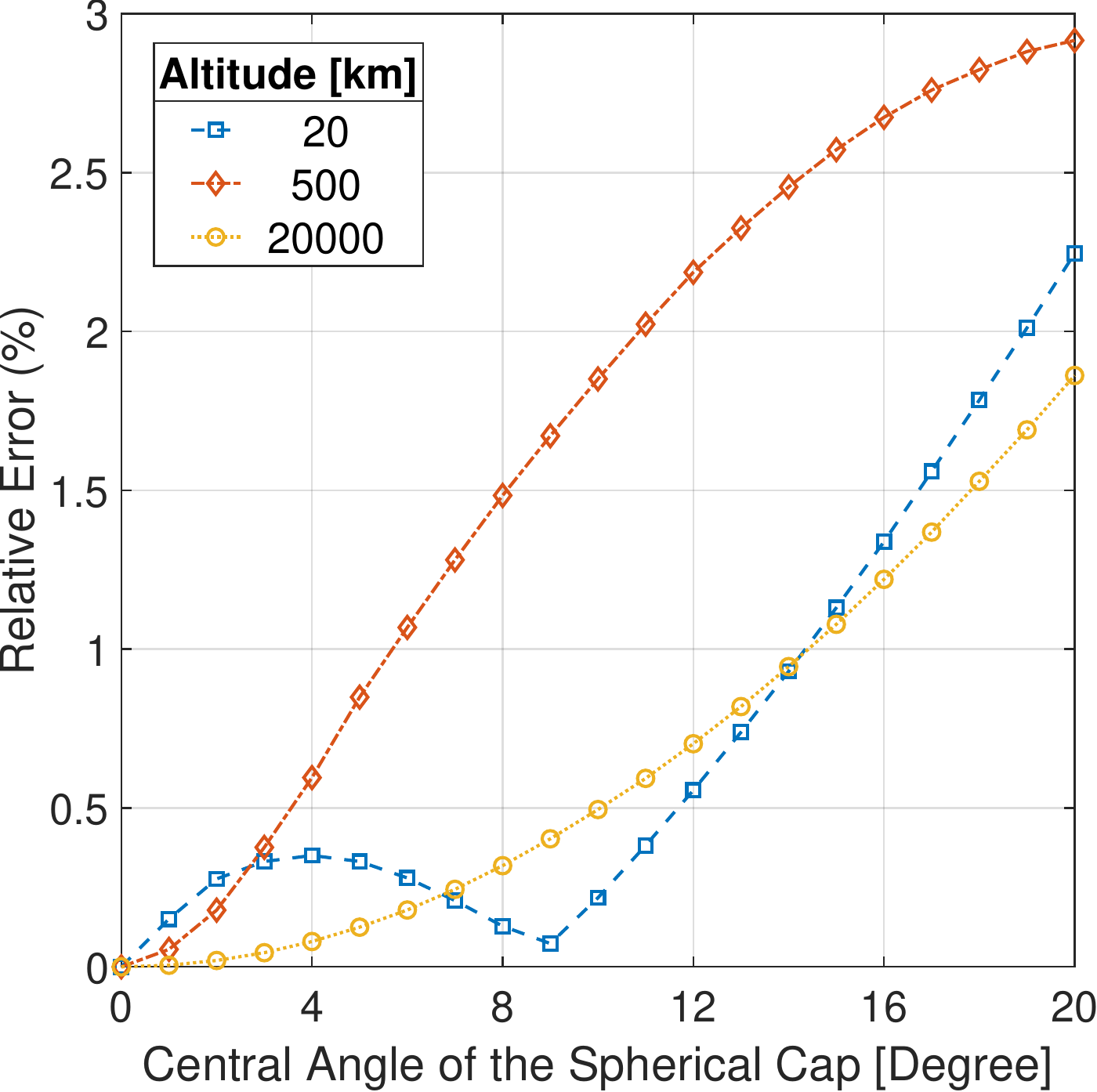}
\caption{The relative error of planar modeling compared to spherical modeling.}
\vspace{-0.2cm}
\label{figure3-2}
\end{figure}

Fig.~\ref{figure3-2} shows the variation in relative error by adjusting the central angle and altitude of the spherical cap. Denote $d_{\mathrm{spherical}}^{(n)}$ and $d_{\mathrm{planar}}^{(n)}$ as the Euclidean distances from the $n^{th}$ point in the spherical BPP and the planar BPP to the typical ground user, respectively, where the ground user is located on the line connecting the zenith to the Earth center. Then, the relative error is defined as
\begin{equation}
    \overline{E} = \frac{1}{N_{\mathrm{NTP}}} \sum_{n=1}^{N_{\mathrm{NTP}}} \frac{\left| d_{\mathrm{spherical}}^{(n)} - d_{\mathrm{planar}}^{(n)} \right|}{d_{\mathrm{spherical}}^{(n)}} \times 100\%,
\end{equation}
where $\left| \cdot \right|$ represents taking the absolute value. {\color{black} The relative error is chosen as the metric to measure the similarity between the two BPPs because the Euclidean distance from NTPs to the user directly influences the performance analysis of the typical user.} The altitudes of three devices, namely HAP, LEO satellite, and MEO satellite, are used as the typical altitudes of the spherical caps, as indicated by the labels in Fig.~\ref{figure3-2}. For improving fairness in comparison, the altitude of the planar BPP $h_{\mathrm{planar}}$ is chosen as the value that minimizes $\overline{E}$ within range $R_{\mathrm{NTP}} \cos\theta_c < h_{\mathrm{planar}} < R_{\mathrm{NTP}}$.

\par
As shown in Fig.~\ref{figure3-2}, the relative error does not have an intuitional relationship with altitude. Modeling LEO satellites on a plane introduces larger relative errors compared to the other two. As the central angle of the spherical cap increases, the relative errors for both types of satellites monotonically increase, while the trend of relative error for HAP is more complex. Next, we provide an example to specifically explain the information within the curves. Assume that $\overline{E} < 0.5\%$ is acceptable. For the central angles of spherical caps $2\theta_c$ corresponding to the line-of-sight (LoS) range of the user, they are $2\theta_c=4.5^{\circ}$ for HAP, $2\theta_c=22^{\circ}$ for LEO satellites, and $2\theta_c=76^{\circ}$ for MEO satellites. Without a doubt, HAP within the LoS range can be approximated as a planar BPP, since 
$4.5^{\circ}<11.8^{\circ}$, where $11.8^{\circ}$ corresponds to the central angle of $\overline{E}=0.5\%$. For satellites, when the user is equipped with a directional narrow-beam antenna for reception, the distribution of satellites within the beam coverage can be approximated by a planar model. When $\overline{E}=0.5\%$, $2\theta_c=3.6^{\circ}$ for LEO satellites and $2\theta_c=9.9^{\circ}$ for MEO satellites. Therefore, the corresponding zenith angle of the directional receiving beam should be less than $83^{\circ}$ for LEO satellites and $26^{\circ}$ for MEO satellites. The conversion from central angle to zenith angle, specifically how central angle $3.6^{\circ}$ relates to zenith angle $83^{\circ}$ for LEO satellites, is given by (\ref{centraltozenith}).

% thetac2 = 83; %波束中心角
% a_ang = 180 - thetac2/2;
% R = (6371 + 500) / sin(a_ang/180*pi);
% b_ang = asin(6371 / R) /pi*180;
% thetac2/2 - b_ang

\subsection{Summary and Discussion}
The topological analysis is the primary difference between the spherical SG and planar SG analytical frameworks. Specifically, the differences between the two are as follows.
\begin{itemize}
    \item NTPs take beam alignment into account, and the NTP-user link may be blocked by the Earth, making the availability analysis of NTPs different from that of terrestrial devices.
    \item When the NTN is modeled as a multi-tier heterogeneous network, or in multiple hop scenarios, the communication association strategy is more complex compared to terrestrial networks.
    \item Unlike ground-based topology which usually uses Euclidean distance to measure the relative position between the user and the communication device, using central angle and zenith angle is more convenient for NTNs.
\end{itemize}
Additionally, we designed a case study to discuss the necessity of spherical modeling.
\begin{itemize}
    \item For NTN, using planar modeling simplifies the derivation, whereas spherical models consider the Earth's curvature, making them more accurate.
    \item Modeling HAPs within the LoS range as a planar BPP introduces acceptable errors, whereas satellites within the LoS range require spherical modeling.
    \item If the user is equipped with directional receiving beams, it may be reasonable to approximate the satellites within the beam as a planar BPP. The higher the altitude of satellites, the greater the modeling error, which requires the receiving beam to be more directional.
\end{itemize}

\section{Channel-Based Analysis} \label{section4}
This section starts with the general channel model under the spherical SG framework. Following that, detailed descriptions of channel modeling are provided for different kinds of scenarios in the next subsections. Representative references in different communication scenarios with their channel models are provided in Table~\ref{tableIV-1}.

\begin{table*}[t]
\centering
\caption{Representative references for channel modeling.}
\label{tableIV-1}
\renewcommand{\arraystretch}{1.1}
\begin{tabular}{|c|c|c|c|}
\hline
References & Network/link & Small-scale fading & Additional consideration \\ \hline \hline
\cite{222302} & Space-to-air link & N/A & N/A \\ \hline
\cite{232301,232303} & Space-to-air link & Nakagami-m fading & N/A \\ \hline
\cite{242211,240305} & Inter-satellite link & Pointing error & N/A \\ \hline
\cite{242201} & Inter-satellite link & Nakagami-m fading & N/A \\ \hline
\cite{200102,222104} & Space-to-ground link & SR fading & Rain attenuation \\ \hline
\cite{200104} & Space-to-ground link & Rayleigh fading & Atmospheric absorption \\ \hline
\cite{220101} & Space-to-ground link & Nakagami-m fading &  Directional antenna gain \\ \hline
\cite{222101,220104} & Space-to-ground link & Mixed lognormal Gaussian & LoS probability  \\ \hline
\cite{232309} & Air-to-ground link & Nakagami-m fading & Directional antenna gain  \\ \hline
\cite{222303} & Air-to-ground link & Nakagami-m fading & Atmospheric absorption  \\ \hline
\cite{242302} & Air-to-ground link & Nakagami-m fading & Misaligned antenna attenuation \\ \hline
\cite{232302,232306} & Air-to-ground link & SR fading & Rain attenuation  \\ \hline
\cite{232305} & Space-to-sea link & SR fading & Rain attenuation \\ \hline
\cite{242301} & Space-to-sea link & Rician fading & N/A \\ \hline
\cite{232305} & Air-to-sea link & Rician fading & N/A \\ \hline
\end{tabular}
\end{table*}

\subsection{General Channel Model}
In the general channel model, the signal power $\eta_r$ at the receiver is given as
\begin{equation}
    \eta_r = \eta_t G H W,
\end{equation}
where $\eta_t$ is the transmission power. $G$ denotes the effective channel gain \cite{242201}, which can be mathematically expressed as
\begin{equation}
    G = G_t G_r \left( \frac{c}{4 \pi f_c} \right)^2,
\end{equation}
where $G_t$ and $G_r$ are antenna gain at the transmitter and receiver, respectively. $c$ is the speed of the light, and $f_c$ is the carrier frequency. $H$ denotes the large-scale fading, which is given as
\begin{equation}
    H = \zeta d^{-\alpha},
\end{equation}
where $\zeta$ is the additional attenuation. $d^{-\alpha}$ is the path-loss, where $d$ is the Euclidean distance between the transmitter and receiver. $\alpha$ is the path-loss exponent, and its value ranges between $2$ and $4$ \cite{200104}. Considering that the application scenarios for NTN are usually in remote areas, $\alpha = 2$ is most commonly used according to the free-space propagation model. Finally, $W$ denotes the small-scale fading, which will be discussed in detail in the following parts of this section.

\subsection{Communications in Aerial and Space Networks}
Based on the general channel model, we start with links in the simplest environments, namely the aerial network and space network.

\subsubsection{Space-to-Air Link}
The atmosphere in space-to-air links is thin, with few obstructions, shadowing effects and atmospheric absorption can be ignored. Therefore, the additional attenuation in space-to-air links is $\zeta = 0$. Authors in \cite{222302} considered that LoS propagation plays a dominant role over the space-to-air link, thus there is no small-scale fading. On the contrary, authors in \cite{232301} and \cite{232303} considered that multi-path effects caused by path impairments cannot be ignored. As a result, the small-scale fading $W$ in the space-to-air link follows the Nakagami-m fading. The PDF of Nakagami-m fading can be expressed as follows \cite{wackerly2008mathematical}:
\begin{equation}
    f_W (w) = \frac{m^m g^{m-1}}{\Gamma(m)} e^{-mw},
\end{equation}
where $m$ is called the shape parameter of Nakagami-m fading, and $\Gamma( \cdot )$ is the Gamma function given by
\begin{equation}
    \Gamma(m) = \int_0^{\infty} t^{m-1} e^{-t} \mathrm{d} t.
\end{equation}
The CDF of Nakagami-m fading is given as follows:
\begin{equation}
    F_W (w) = 1 - \frac{ \Gamma_u(m,mw) }{\Gamma(m)},
\end{equation}
where $\Gamma_u(\cdot,\cdot)$ is the upper incomplete Gamma function,
\begin{equation}
    \Gamma_u(m,mw) = \int_{mw}^{\infty} t^{m-1} e^{-t} \mathrm{d} t.
\end{equation}
Nakagami-m fading can model a wide range of fading environments, from severe fading to light fading, by adjusting the parameter $m$. Therefore, it is widely applied in modeling different types of links.

\subsubsection{Inter-satellite Link (ISL)}
The communication environment of ISLs is similar to that of space-to-air links. Therefore, \cite{242201} uses the same channel modeling as for space-to-air links, where the small-scale fading follows the Nakagami-m distribution. Furthermore, the authors considered the modeling of laser ISL in \cite{242211} and \cite{240305}, and the small-scale fading $W$ follows the pointing error model. The pointing error is caused by the beam deviation during transmission or reception due to the satellites' rapid relative motion and jitter. Given that the deviation angle of the beam is $\theta_d$, the conditional PDF of $W$ is \cite{gappmair2011ook}:
\begin{equation}
\label{pointing_error}
    f_{W \, | \, \theta_d} \left ( w \right ) = \frac{\eta_s^2 w^{\eta_s^2-1 } \cos \theta_d }{A_0^{\eta_s^2}},  \ 0 \leq w \leq A_0,
\end{equation}
where $A_0$ is the fraction of the collected power \cite{farid2007outage}, and $\eta_s$ is the normalized beam waist \cite{ata2022performance}. The deviation angle $\theta_d$ in (\ref{pointing_error}) follows the Rayleigh distribution with variance $\varsigma^2$ \cite{ata2022performance}, 
\begin{equation}
    f_{\theta_d}\left ( \theta \right ) = \frac{\theta}{\varsigma^2}\exp\left ( -\frac{\theta^2}{2\varsigma^2} \right ), \ \theta \geq 0.
\end{equation}

\subsection{Space/Air-to-Ground Communication}
Compared to communications in aerial and space networks, space/air-to-ground Communication involves a more complex environment, thus requiring more detailed modeling.

\subsubsection{Small-Scale Fading}
In space/air-to-ground communication, the impact of multi-path effects is much greater. Nakagami-m fading is the most widely adopted small-scale fading model for the air-to-ground link, while various small-scale fading models are employed for the space-to-ground link. Among them, shadowed-Rician (SR) fading, Nakagami-m fading, Rayleigh fading, and the mixed lognormal Gaussian are some relatively common fading models. It is worth mentioning that SR fading is regarded as the most accurate small-scale fading model for describing space-to-ground links \cite{200102}. It effectively characterizes both the shadowing and multi-path effects of the link. The CDF of the SR fading can be expressed as follows \cite{abdi2003new}:
\begin{equation}
\begin{split}
    F_W (w) & = \left( \frac{2b_0 m}{2b_0 m + \Omega} \right)^m \sum_{z=0}^{\infty} \frac{(m)_z}{z! \, \Gamma(z+1)} \\
    & \times \left( \frac{\Omega}{2b_0 m + \Omega} \right)^z \Gamma_l \left( z+1 , \frac{w}{2b_0} \right),
\end{split}
\end{equation}
where $\Omega$ is the average power of the LoS component, $2b_0$ denotes the average power of the multi-path component excluding the LoS component, and $m$ is known as the Nakagami parameter. $\Gamma(z+1)$ represents the Gamma function, and $(m)_z$ is the Pochhammer symbol. $\Gamma_l \left( \cdot , \cdot \right)$ is the lower incomplete Gamma function, which can be expressed as
\begin{equation}
    \Gamma_l \left( z+1 , \frac{w}{2b_0} \right) = \int_0^{\frac{w}{2b_0}} t^{z} \, e^{-t} \mathrm{d} t.
\end{equation}
The PDF of the SR fading is:
\begin{equation}
\begin{split}
    f_W (w) & = \left( \frac{2b_0 m}{2b_0 m + \Omega} \right)^m 
    \frac{e^{-\frac{w}{2b_0}}}{2b_0}  \\
    & \times _1 \!\! F_1 \left( m, 1 , \frac{\Omega w}{2b_0 \left(2b_0 m + \Omega \right)} \right),
\end{split}
\end{equation}
where $_1 \! F_1 \left( \cdot , \cdot , \cdot \right)$ is the Confluent Hypergeometric function \cite{magnus1967formulas}. Because  processing the PDF $f_W (w)$ is quite challenging, authors in \cite{jia2021uplink} approximated it as
\begin{equation}
    f_W (w) \approx \frac{1}{m_2^{m_1} \Gamma \left( m_1 \right)} w^{m_1 - 1} e^{-\frac{w}{m_2} },
\end{equation}
where $\Gamma (\cdot)$ denotes the Gamma function. $m_1$ represents the shape parameter
\begin{equation}
    m_1 = \frac{m \left( 2b_0 + \Omega \right)^2}{4 m b_0^2 + 4 m b_0 \Omega + \Omega^2},
\end{equation}
and $m_2$ represents the scale parameter
\begin{equation}
    m_2 = \frac{4 m b_0^2 + 4 m b_0 \Omega + \Omega^2}{m \left( 2b_0 + \Omega \right)}.
\end{equation}

\subsubsection{Large-Scale Fading}
As for additional attenuation in large-scale fading for space/air-to-ground links, many studies consider atmospheric absorption \cite{222303,200104}, with rain attenuation being one of the most significant contributing factors \cite{232302,232306}. In addition, some studies also consider the additional attenuation caused by non-line-of-sight (NLoS) transmission \cite{220104}. As a result, the mixed lognormal Gaussian distribution is proposed \cite{222101}:
\begin{equation}
\begin{split}
    \zeta W [ \mathrm{dB} ] & \sim p_{\mathrm{LoS}} \left( \theta_c \right) \mathcal{N} \left( -\mu_{\mathrm{LoS}}, \sigma_{\mathrm{LoS}}^2 \right)\\
    & + p_{\mathrm{NLoS}} \left( \theta_c \right) \mathcal{N} \left( -\mu_{\mathrm{NLoS}}, \sigma_{\mathrm{NLoS}}^2 \right),
\end{split}
\end{equation}
where $\mathcal{N} \left( -\mu, \sigma^2 \right)$ represents Gaussian random variable with mean $-\mu$ and variance $\sigma^2$. $\mu_{\mathrm{LoS}}$, $\sigma_{\mathrm{LoS}}$, $\mu_{\mathrm{NLoS}}$, and $\sigma_{\mathrm{NLoS}}$ depend on the propagation environment \cite{al2020modeling}. The probability of establishing LoS links between NTP and ground users is
\begin{equation}
    p_{\mathrm{LoS}} \left( \theta_c \right) = \exp \left( - \frac{R_{\mathrm{NTP}} \beta \sin\theta_c}{R_{\mathrm{NTP}}\cos\theta_c - R_{\oplus}} \right),
\end{equation}
where $\beta$ also depends on the propagation environment given in \cite{al2020modeling}, $\theta_c$ is the central angle between NTP and user, $R_{\oplus}$ is radius of the Earth, and $R_{\mathrm{NTP}}$ is radius of the sphere where NTPs are located. The probability of establishing NLoS links is $p_{\mathrm{NLoS}} \left( \theta_c \right) = 1 - p_{\mathrm{LoS}} \left( \theta_c \right)$. Note that the mixed lognormal Gaussian distribution describes the distribution of the product of additional attenuation $\zeta$ and small-scale fading $W$. In the mixed lognormal Gaussian distribution,  $W$ follows Gaussian distribution $\mathcal{N} \left( -\mu_{\mathrm{LoS}}, \sigma_{\mathrm{LoS}}^2 \right)$ given that the link is LoS.

\subsubsection{Antenna Beam}
Due to the long communication distance between NTP and ground users, NTP is equipped with directional antennas, which use narrower beam widths to achieve higher antenna gain. Next, we present the beam alignment pattern and beam modeling under the spherical SG framework. 

\par
Overall, there are four types of beam alignment strategies. Some studies suggested that satellites align the beam center with the sub-satellite point, which is the intersection of the line connecting the satellite and the Earth center with the Earth surface \cite{200104,210105}. The second alignment strategy is that satellites align the beam center directly with the typical user \cite{200102}. Furthermore, a satellite is equipped with multiple beams, and beams are pointed towards multiple fixed cells \cite{jia2021uplink, na2021performance}. Although these two articles only considered modeling for a single satellite, thus belonging to the SG analytical framework rather than the spherical SG, the alignment strategy for multiple fixed beams is still worth mentioning. Finally, satellites are assumed to be equipped with multiple flexible spot beams in \cite{242109}. This beam alignment strategy has been proven to effectively reduce inter-beam interference. A potential research direction for spherical SG is interference and coverage analysis under a bidirectional alignment model, which can be viewed as an extension of the spot beam model-based analysis. In this approach, the satellite transmits using directional spot beams, while users receive directionally to minimize inter-beam interference. 

\par
There are also three ways to describe the gain loss due to beam misalignment. Firstly, authors in \cite{242302} introduced the attenuation factor to account for misaligned antenna attenuation, which can be regarded as a type of additional attenuation $\zeta$. Secondly, the misaligned antenna attenuation is modeled as small-scale fading in \cite{240305}. The distribution of attenuation, known as the pointing error, was provided in the previous subsection. Finally, and most commonly, misaligned antenna attenuation is modeled by the antenna gain $G$. $G$ is expressed as a function of the deflection angle $\varphi$ \cite{220101}, where the deflection angle $\varphi$ is the angle between the beam alignment direction and the direction from the NTP to the user. The Flat-top model is one of the most commonly applied antenna gain models \cite{balanis2015antenna}:
\begin{equation}
    G(\varphi) = \left\{\begin{matrix} G_m,  & \left | \varphi \right | \leq  \varphi_{\mathrm{3dB}}, \\  0,    &  { \mathrm{otherwise}}, \end{matrix}\right.
\end{equation}
where $G_m$ is the maximum gain, $\varphi_{\mathrm 3dB}$ denotes the half-power beamwidth. Furthermore, another practical directional beam gain model is called the Gaussian antenna pattern \cite{gagliardi2012satellite}: 
\begin{equation}
    G(\varphi) = G_m \, 2^{-\varphi^2 / \varphi_{\mathrm{3dB}} ^2}.
\end{equation}

\subsection{Space-Air-Ground(-Sea) Integrated Network}
This subsection explores the channel models and communication modes under two complex network structures, namely space-air-ground integrated network (SAGIN) and space-air-ground-sea integrated network (SAGSIN). 

\subsubsection{Space/Air-to-Sea Link}
First is the channel modeling for space-to-sea and air-to-sea links. In \cite{232305}, authors considered the space-to-sea link using the same channel model as the space-to-ground link. In this model, small-scale fading follows SR fading and rain attenuation is accounted for as additional attenuation. Furthermore, the authors in this study introduced, for the first time, an air-to-sea link channel model suitable for the SG analytical framework, which was later adopted in \cite{242301}. Rician fading is considered as the small-scale fading for the air-to-sea link, describing one strong direct LoS component and many random weaker components over the sea. The CDF of Rician fading can be expressed as \cite{abdi2001estimation}:
\begin{equation}
    F_W (w) = 1 - \int_{\sqrt{2K+2}}^{\infty} t \exp\left( -\frac{t^2 + 2K}{2} \right) I_0 \left( \sqrt{2K} t \right) \mathrm{d} t,
\end{equation}
where $K_{\mathrm{Rician}}$ denotes the Rician $K$ factor, and $I_0 (\cdot)$ is the modified Bessel function of the first kind. Moreover, the PDF of the Rician fading is
\begin{equation}
    f_W (w) = (2K+2) \,w \, I_0 \left( 2w \sqrt{K^2 + 2K} \right) e^{-(K+1)w^2 - K}.
\end{equation}

\subsubsection{Relay Communication}
Next, we show several relay communication scenarios and their motivations from the perspective of the channel fading. In the first scenario, satellite gateways are introduced as relays in space-to-ground communications. Although satellite gateways and ground users experience similar propagation conditions when communicating with satellites, the internal noise power of the gateway is lower \cite{200102}. After receiving the signal from the satellite, the gateway demodulates and forwards it to the user, significantly expanding the coverage area and increasing the coverage probability. Another study proposed that in a cell-free network, HAPs act as relays for communication between satellites and ground terminals \cite{232306}. The study highlighted that the satellite-to-ground link experiences long distances and complex communication environments. In contrast, the satellite-to-HAP link has a simpler communication environment, and the HAP-to-ground link covers a shorter distance. Consequently, both links are more favorable than the satellite-to-ground link. Furthermore, the article demonstrated through numerical results that introducing HAPs as relays can effectively improve coverage performance. 

\par
By incorporating satellites as relays, the authors in \cite{242301} improved the coverage performance of the marine link between ship and shore base station. With common parameter settings, the transmission success probability experiences a multi-fold increase. This improvement is attributed to the space link's wider bandwidth and greater transmission power compared to the maritime link. In the end, the article \cite{242211} considered a scenario involving uplink transmission from IoT devices in remote areas. For IoT devices with limited transmission power, directly accessing the network through GEO satellites is challenging. Because the distance between IoT devices and LEO satellites is only one-tenth of that between GEO satellites, IoT devices can indirectly achieve long-distance communication with GEO satellites by using LEO satellites as relays. 

\par
{\color{black}It is worth mentioning that existing research primarily focuses on describing the necessity and advantages of introducing relays while neglecting their costs and disadvantages. First of all, the introduction of relays increases energy consumption in communication, and this additional energy cost is often borne by NTPs, which typically have higher energy costs compared to terrestrial devices. A potential solution is to compare the energy efficiency of relay-based communication with that of non-relay communication. Energy efficiency takes into account the data rate gains achieved through the introduction of relay communication, as well as the costs associated with the additional energy consumption, and analyzes whether the incorporation of NTP relays can provide communication advantages. Furthermore, the introduction of relays adds technical complexity to communication. In scenarios like long-distance communication, where precise beam alignment is required, adding relays means more pairs of transmitters and receivers are needed. The number of hops can serve as a metric for the added technical complexity and the extra cost of equipment. The combination of spherical SG and graph theory may help in estimating the number of hops.}

\subsection{Summary and Discussion}
Considering that the current spherical SG framework is slightly different from the planar SG framework in terms of channel-based metric analysis, we mainly focus on channel modeling in this subsection. The primary conclusions are given as follows.
\begin{itemize}
    \item The channel gain is mathematically represented as the product of antenna gain, large-scale fading, and small-scale fading. Large-scale fading primarily includes path loss, atmospheric absorption (rain attenuation), and additional losses in NLoS conditions.
    \item SR fading is considered the most accurate small-scale fading model for space-to-ground links. Rician fading is used to model small-scale fading in space/air-to-sea links. When LoS links dominate, Nakagami-m fading can be widely applied to various environments, making it commonly used across different types of links. 
    \item Directional antenna gain is mentioned in some studies, while most research applies the flat-top gain model for the sake of simplifying the model. In addition, some studies consider a highly aligned spot beam model in long-distance satellite communication, where pointing error or misaligned antenna attenuation becomes an essential factor to consider.
    \item Finally, from the perspective of channel fading, we explain the motivation for applying LEO satellites, satellite gateways, and ships as relays in a spherical SG framework.
\end{itemize}

\section{Advanced Topics}
This section discusses several advanced topics in NTN: routing, satellite clusters, security issues, energy harvesting, and satellite-enabled positioning problems. Unlike the previous section, which focused more on single-hop link layer analysis, this section discusses application scenarios under the spherical SG framework.

\subsection{Routing in NTN} \label{section5-1}
\subsubsection{Motivation}
The routing design in NTN networks is an issue of significant concern. For long-distance routing, NTPs have an irreplaceable advantage. Due to their higher altitude and larger LoS range, NTPs can achieve routing with fewer hops and shorter paths. 

\par
Current research on NTN routing can be mainly categorized into two types: graph theory-based routing and stochastic algorithm-based routing \cite{240306}. NTN routing studies typically involve the design of optimal routing algorithms and the performance evaluation of these algorithms. On the one hand, graph theory-based routing is designed according to predetermined start and end positions \cite{9348676_1,shen2020dynamic_2} or even deterministic network topologies \cite{knight2011internet,geng2021agent_3}. However, considering that the topology of NTN is dynamic, graph theory-based routing has significant limitations. On the other hand, stochastic algorithm-based routing is suitable for dynamic routing design \cite{rabjerg2021exploiting}. However, the algorithms applied in routing, such as the particle swarm optimization algorithm \cite{8068282_5} and the ant colony algorithm \cite{zhao2021multi_4}, exhibit uncertainty in routing performance. As a result, the algorithms' performance cannot be analytically expressed and can only be estimated through multiple rounds of high-computational complexity simulations. In summary, neither graph theory-based routing nor stochastic algorithm-based routing can perfectly address dynamic routing design and low-complexity performance evaluation simultaneously.

\par
Based on the above requirements, SG-based routing is proposed. Since the positions of NTPs follow a specific stochastic distribution, the routing design under the SG framework is built on dynamic topologies. Additionally, the routing performance can be estimated using analytical expressions, resulting in low computational complexity for performance evaluation.

\subsubsection{Research Foundation}
{\color{black}The basic SG-based routing framework can be divided into three steps. Firstly, model the NTP according to a specific distribution, such as the homogeneous spherical BPP. Secondly, select a group of NTP devices as routing relays in the point process according to a certain strategy, which will be discussed in the next paragraphs of this part. Finally, since the spatial distribution of NTP follows a specific SG model, the performance of the routing is generally analyzable. We need to derive the analytical expressions of routing performance, such as latency, based on specific routing strategies.} So far, spherical SG-based routing strategies are classified into two types. 

\par
\begin{figure}[ht]
\centering
\vspace{-0.2cm}
\includegraphics[width = \linewidth]{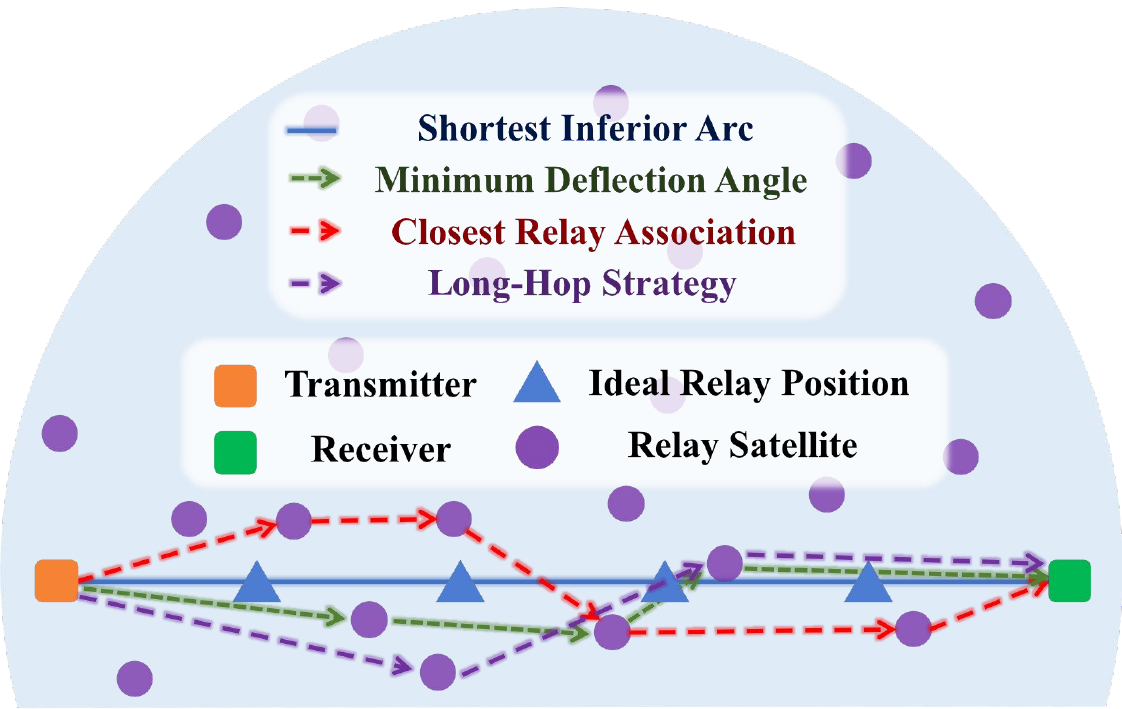}
\caption{Typical relay selection strategies in multi-hop routing.}
\vspace{-0.2cm}
\label{figure5-1}
\end{figure}

The first type extends from ground SG-based routing strategies. Fig.~\ref{figure5-1} presents two typical routing strategies, called the minimum deflection angle strategy and the long-hop strategy. Denote the angle between the routing direction and the shortest inferior arc connecting the transmitter (route start point) and receiver (route endpoint) \cite{220302} as the deflection angle. In the minimum deflection angle strategy, each NTP selects the next relay with the smallest deflection angle relative to itself, aiming to route along the shortest path. As shown in Fig.~\ref{figure5-1}, the long-hop strategy selects the NTP nearest to the receiver within a reliable communication range as a relay \cite{240306}. The reliable communication range is a finite sector-shaped area designed to facilitate routing along a specified direction, reducing the probability of communication interruption due to excessively long single-hop distances \cite{haenggi2005adhoc}. Therefore, authors in \cite{haenggi2005routing} also proposed the short-hop strategy for comparison, where a device chooses its nearest device as the next relay. {\color{black}In the first type of routing strategy, each NTP relies only on local information in relay selection, which is unsuitable for spherical routing.} Taking routing in LEO satellite constellations as an example, it typically requires only a few hops to transmit the signal to the receiver.

\par
As a result, the second type of routing strategy is proposed by utilizing global information and determining the required number of hops at the beginning of the routing \cite{wang2025satellite}. Authors in \cite{220302} proposed the closest relay association strategy shown in Fig.~\ref{figure5-1}. The optimal number of hops and ideal relay positions are first determined by optimization. The optimization objective in \cite{220302} is minimizing the propagation latency, which is the ratio of the transmission distance to the speed of light. Then NTPs closest to these ideal relay positions are selected in the route. Since the estimate of propagation latency does not involve channel modeling, the spherical SG framework has not been fully developed. Therefore, authors in \cite{240305} further introduce the inter-satellite channel model and optimize the transmission latency, coverage probability, and availability. Considering the difficulty of multi-objective optimization, the authors used transmission latency as the optimization objective and the other two metrics as constraints. The ideal relay positions are obtained through optimization, and then the routing is designed according to the closest relay association strategy. Finally, both \cite{220302} and \cite{240305} indicate that, due to the consideration of global information, the second type of routing strategy performs better than the first type in terms of performance metrics, such as latency, coverage probability, and availability probability.

\subsubsection{Future Research Direction}
The literature on routing based on spherical SG is still limited, and many routing strategies in more practical scenarios have not yet been studied. Firstly, expanding single-path routing to multi-path routing is a potential research direction. Multi-path routing increases the reliability and robustness of routing by introducing redundancy and additional network load. Even if a link is interrupted during transmission, the signal can still potentially reach the receiver through other paths. Introducing concepts from graph theory and mesh network topologies \cite{chaudhry2020free} may offer insights for implementing multi-path routing under the SG framework.

\par
\begin{figure*}[ht]
\centering
\vspace{-0.2cm}
\includegraphics[width = \linewidth]{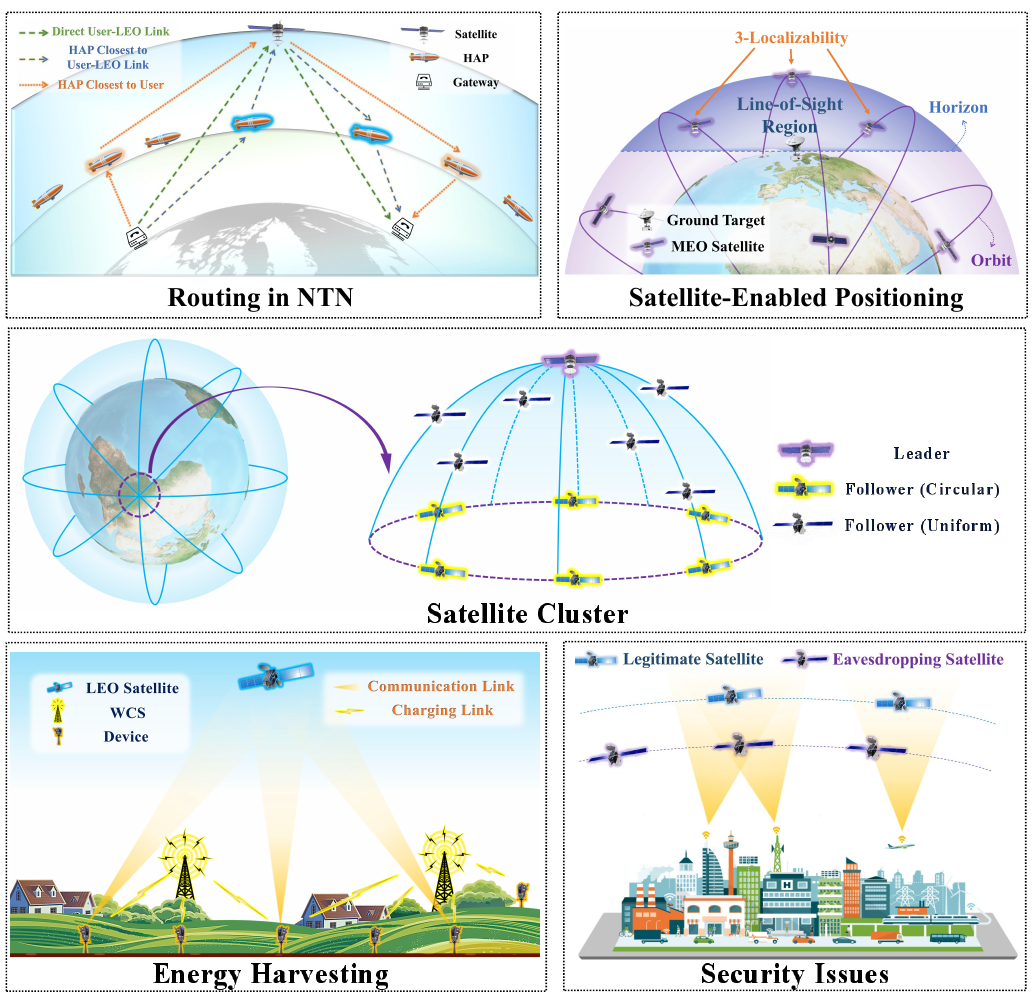}
\caption{Schematic diagram for advanced topics.}
\vspace{-0.2cm}
\label{figure5-2}
\end{figure*}

Secondly, studies have only considered inter-satellite routing in spherical SG. With the expansion of routing scenarios from single-tier networks to multi-tier networks, more types of NTPs can participate in routing. {\color{black}This makes the routing design more complex but also more realistic. For example, many routing requests are initiated from the ground and eventually return to the ground. Additionally, some LEO satellite constellations like OneWeb and Globalstar do not configure ISLs \cite{del2019technical}. In this case, designing a hybrid relay routing system involving ground gateways and LEO satellites is necessary. Then, routing in a SAGIN is taken as an example to illustrate the complexity of cross-tier routing, as shown in Fig.~\ref{figure5-2}. Assuming that the locations of the gateways are fixed, we need to find one LEO satellite and up to two HAPs as relays. Such a simple routing topology includes three types of channel links: air-to-ground, air-to-space, and space-to-ground. Furthermore, Fig.~\ref{figure5-2} presents three HAP relay selection strategies, which can be referred to \cite{242211}. It can be seen that the decision to select relay HAPs and the positions of these HAP relays have a significant relationship to the position of the satellite, making the routing design more difficult. In conclusion, routing design in multi-tier heterogeneous networks is a challenging but meaningful research direction.}

\subsection{Security Issues} \label{section5-2}
\subsubsection{Motivation}
With the extensive deployment of NTN, the widespread NTPs in the air and space also bring potential eavesdropping and jamming threats \cite{yue2023low}. NTPs may intercept communication data from the ground to the air or space, posing an eavesdropping risk. Due to the propagation of electromagnetic waves and interference from other devices, communication links in NTN may experience malicious interference from other NTPs, affecting communication quality. Furthermore, due to the dynamic nature of NTN, atmospheric conditions, and multi-path fading, the signal quality is subject to frequent changes \cite{240304}. At times, the main channel may experience poor conditions, while an eavesdropper may receive a higher quality signal than the main channel, exacerbating the security threat.

\subsubsection{Research Foundation}
In the planar SG framework, some works analyze the performance of secure communication. The work in \cite{6512533} pioneered the application of physical layer security (PLS) in cellular networks. Later, the study in \cite{9627722} extended this concept to millimeter-wave relaying networks. Additionally, the research presented in \cite{7906491} and \cite{7839915} incorporated artificial noise (AN) as a typical PLS technique into the secure transmission. Building on the foundation of terrestrial networks, the authors in \cite{220301} first introduced the scenario of satellites acting as eavesdroppers within the spherical SG framework. The eavesdropping satellites are modeled as a homogeneous BPP, while the relative position of ground terminals and the legitimate satellite is considered deterministic. Based on this modeling, the authors derived the probability of not being eavesdropped by any satellite under different beam alignment strategies in uplink communication. Furthermore, authors in \cite{230303} considered the security of downlink communication in LEO satellite networks. Unlike \cite{220301}, the authors in \cite{230303} not only modeled the ground-based eavesdroppers as a spherical PPP but also assumed the legitimate satellite constellation followed another spherical independent spherical PPP. Finally, the authors in \cite{240303} established multi-tier satellite networks, assuming one tier as the legitimate satellite constellation and the others as potential eavesdroppers. The AN technique was applied to reduce the risk of eavesdropping. Notably, this is currently the only study that applies PLS techniques within the spherical SG framework.

\subsubsection{Future Research Direction}
Here, we give three potential research directions. Firstly, in addition to eavesdropping, jamming can also pose security issues. Jammers transmit interfering signals on the same frequency as the target communication to degrade the quality of the receiver. In scenarios where NTPs act as jammers, it is meaningful to establish an analytical framework for the communication performance of the receiver. Based on this framework, the improvements in NTN performance brought by anti-jamming techniques can be quantified with low complexity. The challenges in Secondly, beyond the space-to-ground link, exploring the security issues in other links mentioned in Sec.~\ref{section4} also holds significant research value. Thirdly, in addition to AN, we can also establish performance analysis frameworks for other PLS techniques, such as frequency hopping and beamforming technology. These three types of techniques reduce the eavesdroppers' received SINRs by respectively decreasing the transmission power, bandwidth, and coverage area. {\color{black}The challenges faced by the three potential research directions mentioned above primarily lie in mathematical modeling and formulation, as well as the trade-offs between techniques. Developing accurate models for PLS techniques like frequency hopping requires a comprehensive understanding of operating principles and the environments in which they are deployed. These contents extend beyond the traditional mathematical framework of spherical SG. Studying security issues under the spherical SG framework can be mathematically intensive and may require extensive simulations. In addition, analyzing the performance of PLS techniques often involves trade-offs. For example, increasing the frequency hopping rate can enhance security but might reduce the overall data rate. Balancing these trade-offs while optimizing for security is also a significant challenge.
}

\subsection{Satellite Cluster} \label{section5-3}
\subsubsection{Motivation}
With the construction of mega-constellations, interference between closely spaced small satellites has become an issue that cannot be ignored. Inspired by this, the concept of satellite clusters is proposed, aiming to enhance communication performance through coordinated transmission and reception of signals among a group of physically adjacent satellites \cite{liu2018survey}. A satellite cluster consists of one leader and multiple follower satellites, which exchange information through ISLs. Specifically, satellite clusters can be categorized into two types based on the hardware structure between the leader and follower satellites. 

\par
The first type is a temporary cluster formed by several physically adjacent satellites from the same constellation \cite{242201}, and therefore, these satellites have similar hardware configurations. Please note that the distance between them is relatively close, with the actual distance potentially reaching tens or even hundreds of kilometers. This means that the channels from the satellites in the cluster to the terminals may not be highly correlated, allowing for the realization of spatial diversity \cite{yu2016virtual}. Spatial diversity can significantly enhance the communication quality of the cluster compared to a single satellite. For example, LEO satellites do not always have an LoS link with ground users. In fact, the probability of establishing an LoS link between the satellite and the user depends on their elevation angle, and spatial diversity ensures that the event of each satellite establishing an LoS link with the ground user is independent \cite{al2020modeling}. The leader can select a follower with an LoS link as the serving satellite to avoid performance degradation caused by obstructions.

\par
The second type of satellite cluster is constructed by making slight changes to the orbital configuration, causing the follower satellites to orbit around the leader satellite over the long term \cite{popov2021development}. Because the satellites in the cluster are fixed, this helps establish stable communication links between the leader and follower satellites. Additionally, the functionality of the follower satellites can be simplified. For example, they can be used solely for coverage extension and throughput enhancement, without needing a full protocol stack \cite{232203}. As a result, the hardware configuration of the follower satellites can be less complex than that of the leader satellite, making it easier to expand the cluster scale.

\subsubsection{Research Foundation}
Currently, there are two studies in the field of spherical SG that discuss the structure of satellite clusters \cite{232203,242201}. As shown in Fig.~\ref{figure5-2}, authors in \cite{232203} introduced two kinds of clusters: circular clusters and uniform clusters. A circular cluster refers to follower satellites arranged at equal intervals along the edge of a spherical cap centered on the leader satellite \cite{eyer2007formation}. The leader satellites in a constellation with circular clusters are modeled as a homogeneous BPP, while follower satellites are deterministically distributed along the circle. Since the follower satellites are equidistant from each other, the antenna alignment for ISL communication is expected to be simpler. 

\par
As for the uniform cluster, the distribution of follower satellites forms a homogeneous BPP distributed on a spherical cap centered on the leader satellite. The uniform cluster effectively utilizes the entire spherical cap, making it more suitable for dense deployments. Based on the uniform cluster modeling, authors in \cite{242201} investigated the coverage probability of joint transmission among satellites in a cluster. They defined the signal-to-interference ratio as the ratio of the sum of signal power from all satellites within the cluster to the sum of interference power from the remaining satellites outside the cluster.

\subsubsection{Future Research Direction}
The only existing technical study focuses on the coverage probability performance gain provided by a uniform cluster structure when the hardware configuration is the same. In the same scenario, analyzing the coverage resilience is also an interesting topic. When a follower satellite malfunctions or temporarily ceases operation due to full load, its coverage area can be serviced by other satellites in the cluster. Secondly, the communication performance gain when the leader and follower satellites have different hardware configurations is worth exploring. In a cluster, follower satellites can transmit information to each other via high-quality ISLs with short distances and minimal channel disturbances. By using follower satellites as relays, the channel capacity and energy efficiency can be enhanced through parallel transmissions by multiple follower satellites. Lastly, the risk assessment between circular clusters and uniform clusters is another aspect to consider. Based on the spherical SG framework, we can analyze the average distance between any satellite and its nearest neighboring satellite as a measure of security. {\color{black}Compared to the analysis of non-clustered LEO satellite networks, the challenges in analyzing satellite followers primarily lie at the topological level. The communication/security distance analysis between satellites and the ground, between leader and follower satellites, and among satellite followers is significantly more complex than the contact distance mentioned above.}

\subsection{Energy Harvesting} \label{section5-4}
\subsubsection{Motivation}
Due to low communication traffic and high costs of deploying terrestrial BSs and fiber optics, many remote areas lack terrestrial communication infrastructure. In these under-served locations, the advancements in NTN communications are offering a promising solution through aerial access \cite{araniti2021toward}. At the same time, it is challenging to regularly charge or replace batteries for terrestrial devices, such as sensors in remote areas \cite{huang2019wireless}. In order to achieve the vision of self-sustainability, apart from reducing energy consumption and relying on solar energy, it is also crucial to provide wireless charging capabilities and develop energy harvesting techniques for such devices \cite{liu2019toward}. Deploying several wireless charging stations (WCS) in remote areas can meet the changing needs of a large number of devices, undoubtedly reducing the cost of energy supply.

\subsubsection{Research Foundation}
The performance analysis of terrestrial energy harvesting systems based on the SG framework has received extensive attention. In \cite{huang2011throughput}, the authors first applied the SG analytical framework to the performance analysis of wireless charging networks, using throughput as the metric for charging efficiency. Furthermore, a more realistic scenario was considered in \cite{flint2014performance}, where the energy harvesting system is activated only when the collected instantaneous energy exceeds the energy required for circuit operation. Based on the assumption of a minimum circuit energy activation threshold, the authors in \cite{kishk2016downlink} conducted an overall performance evaluation of a hybrid system that includes an energy harvesting subsystem and a communication subsystem using the SG framework. Specifically, they defined a time-slot that can be partitioned into a charging sub-slot and a communication sub-slot. By adjusting the proportions of these two sub-slots, the network throughput was optimized. So far, \cite{240303} is the only literature under the spherical SG framework that considers energy harvesting in NTN networks. As shown in Fig.~\ref{figure5-2}, a hybrid system consisting of a terrestrial energy harvesting subsystem and a satellite communication system is considered. Specifically, in addition to relying on WCS for wireless charging, the authors in \cite{240303} also considered the scenario where communication devices are powered by a hybrid energy supply, including batteries.

\subsubsection{Future Research Direction}
Unlike energy harvesting in terrestrial networks, the terrestrial energy harvesting subsystem and the NTN communication subsystem are relatively independent in spatial distribution in the NTN-based wireless-powered scenario. Since both subsystems have been relatively well studied separately, to distinguish the performance evaluation of the hybrid system from the combination of the two subsystems, the research focus of energy harvesting in NTN is expected to be on the interaction between the two systems. An interesting topic is the time allocation issue between the energy harvesting sub-slot and the communication sub-slot. If the communication sub-slot is too long, it limits the time available for energy harvesting, resulting in insufficient transmission power and consequently a lower probability of successful communication and a lower data transmission rate. Conversely, shortening the communication sub-slot reduces the time available for data transmission, making it challenging to complete data packet transmission within the limited time. Unlike the study \cite{kishk2016downlink}, which considered communication and energy harvesting systems as static or quasi-static, {\color{black}the satellite communication system is dynamic relative to the ground, making the analysis more difficult. Specifically, we can design an algorithm that allows the satellite to transmit signals when it is closer to the ground signal transmitter and disconnect communication to continue energy harvesting when the associated satellite is farther away. The performance of this algorithm in terms of communication and energy harvesting can be analyzed using the spherical SG analytical framework.} Based on the analytical results derived by SG, the time allocation of the two sub-slots can be optimized. It is important to note that satellites need to be modeled by a stochastic-orbital or a fixed-orbital one to involve concepts such as the satellite communication window period and to simulate the continuous movement of the satellite.

\subsection{Satellite-Enabled Positioning} \label{section5-5}
\subsubsection{Motivation}
After decades of development, several global navigation satellite systems (GNSSs), such as GPS, Beidou, and Galileo have been put into operation, and satellite-enabled positioning technology has become a relatively mature technology \cite{cai2015precise}. So far, positioning services are primarily carried out by MEO satellites and a small number of GEO satellites. However, relying on existing GNSSs faces challenges in handling the surge in positioning demands from future networks. Besides launching more MEO satellites to expand the capacity of the positioning system, involving the deployment of mega LEO satellite constellations in auxiliary positioning is another potential solution \cite{liao2023integration}. Due to LEO satellites' lower altitude, LEO satellite-enabled positioning systems are expected to achieve lower positioning latency \cite{ferre2022leo}. Given the current trends of expanding constellation size and increasing structure complexity in satellite positioning systems, there is a strong motivation to apply spherical SG to positioning performance analysis.

\subsubsection{Research Fundation}
Several studies have already analyzed the performance of ground-based positioning systems using the SG framework. In \cite{schloemann2015localization}, authors first introduced the concept of $K$-localizability probability and used it as a metric to measure localization performance. Serving as one of the most fundamental and important metrics for localization and an important link between the SG framework and the localization system, the definition of the $K$-localizability probability will be given in the next paragraph. Next, authors in \cite{schloemann2015toward} considered the $K$-localizability probability performance under the condition that base stations used for localization can be scheduled. Furthermore, authors in \cite{christopher2018statistical} derived the distribution of the Cram$\acute{e}$r–Rao lower
bound (CRLB) conditioned on the
number of devices participating in positioning. In other words, \cite{christopher2018statistical} established the relationship between the CRLB and the $K$-localizability probability. So far, NTP-enabled positioning has not been studied under the spherical SG framework. Only the work \cite{230302} studied the passive radar detection performance analysis in the LEO/MEO satellite system, which demonstrates the potential for interaction between NTP-enabled localization and spherical SG.

\subsubsection{Future Research Direction}
{\color{black}The challenges in satellite-enabled positioning research based on spherical SG lie in the current lack of foundational studies. There is a need to establish channel models, performance metrics, and analytical frameworks. First, regarding performance metrics, we propose a potential direction for future research.} $K-$localizability probability is the probability that a ground target can simultaneously detect $K$ NTPs' signals. Successful detection requires the received SINR to reach a predefined threshold. The mathematical definition of $K-$localizability probability is given as follows: 
\begin{equation}
    P^L (K) = \prod_{k=1}^K \mathbbm{P}\left[ \mathrm{SINR}_k > \gamma \right],
\end{equation}
where $\mathrm{SINR}_k$ is the instantaneous received SINR of the $k^{th}$ NTP, and $\gamma$ is the threshold. Furthermore, if the satellite is below the ground target's horizon, we assume the target can not receive the signal and the received SINR is $0$ in this case. An example of $K-$localizability is shown in Fig.~\ref{figure5-2}.

\par
Research indicated that increasing the number of satellites involved in localization can effectively reduce localization errors \cite{specht2015accuracy}. Therefore, the analysis of $K$-localizability probability in NTP-enabled localization systems is expected to be a foundational work in this area. A further potential research direction is the performance analysis of geometric localization algorithms. Current satellite localization relies on geometric positioning techniques, and SG is a mathematical tool well-suited for analyzing geometric topology and channel randomness. This topic encompasses both algorithm design and analytical framework derivation. It is foreseeable that the core of this research direction will include the estimation of the probabilistic fuzzy region, which is also called circular error probable region \cite{lewis2007effects}.

\subsubsection{Case Study}
Based on the previous discussions, we can identify several research gaps within the spherical SG framework. Firstly, the modeling and performance analysis of MEO satellites have yet to be explored. Secondly, the DSBPP model has not been applied. Lastly, within these advanced topics, only satellite-enabled positioning has not been mentioned. Therefore, we have investigated the $K$-localizability probability for DSBPP-modeled MEO satellite networks. In fact, the design of this scenario is not only to fill the research gaps, but the scenario itself is also reasonable. The DSBPP model is suitable for accurately modeling MEO satellites, and most current MEO satellite constellations are used for global positioning.

\par
Based on the parameters of the GPS constellation, each orbit is deployed with $4$ satellites. If the total number of satellites is not a multiple of $4$, we assume there is one orbit with fewer than $4$ satellites. Since the constellation scale is small, the bandwidth required for localization is relatively low, resulting in low difficulty for frequency coordination. Therefore, we assume there is no co-channel interference. The small-scale fading between the MEO satellite and the terrestrial target follows the SR fading. Finally, values of the simulation parameters are provided in Table~\ref{tableV}.

\begin{table}[t]
\centering
\caption{Simulation Parameters.}
\label{tableV}
\resizebox{\linewidth}{!}{
\renewcommand{\arraystretch}{1.1}
\begin{tabular}{|c|c|c|}
\hline
Notation &  Physical meaning  & Value  \\ \hline \hline
$R_{\oplus}$  & Radius of the Earth & $6371$~km  \\ \hline
$N_M$  & Number of satellites on the orbit  & $4$  \\ \hline
$\eta_t$  & Transmission power & $18$~dBW    \\ \hline
$f_c$  & Center frequency & $1.575$~GHz  \\ \hline
$G_t$, $G_r$  & Antenna gain & $20.9$~dBi  \\ \hline
$\zeta$   & Additional attenuation  & $-2$~dB   \\ \hline
$ (\Omega,b_0,m)$ & Parameters of SR fading  & $ (1.29,0.158,19.4)$ \\ \hline
$\sigma^2$  & Noise power  & $-98$~dBm \\ \hline
$\gamma$ &  Coverage threshold & $0$~dB \\ \hline
\end{tabular}
}
\end{table}

\begin{table*}[t]
\centering
{\color{black}
\caption{{\color{black}Roadmap table for advanced topics.}}
\label{tableV-1}
\renewcommand{\arraystretch}{1.1}
\resizebox{\linewidth}{!}{
\begin{tabular}{|c|c|c|c|}
\hline
Topic & Current & Short term & Long term \\ \hline \hline
Routing in NTN & Simple strategy  & Multi-tier and multi-path routing & More routing constraints (congestion) \\ \hline
Security issues & Eavesdropping & PLS and anti-eavesdropping strategies & Broader security issues (anti-jamming) \\ \hline
Satellite cluster & Cluster modeling & Performance analysis of clusters & Cluster-based communication strategy design \\ \hline
Energy harvesting & N/A & Wireless-powered performance analysis & Slot allocation and charging strategies design \\ \hline
Positioning & N/A & Availability and localizability analysis & Positioning errors analysis in geometric localization \\ \hline
\end{tabular}
}}
\end{table*}

\begin{figure}[ht]
\centering
\vspace{-0.2cm}
\includegraphics[width = 0.8\linewidth]{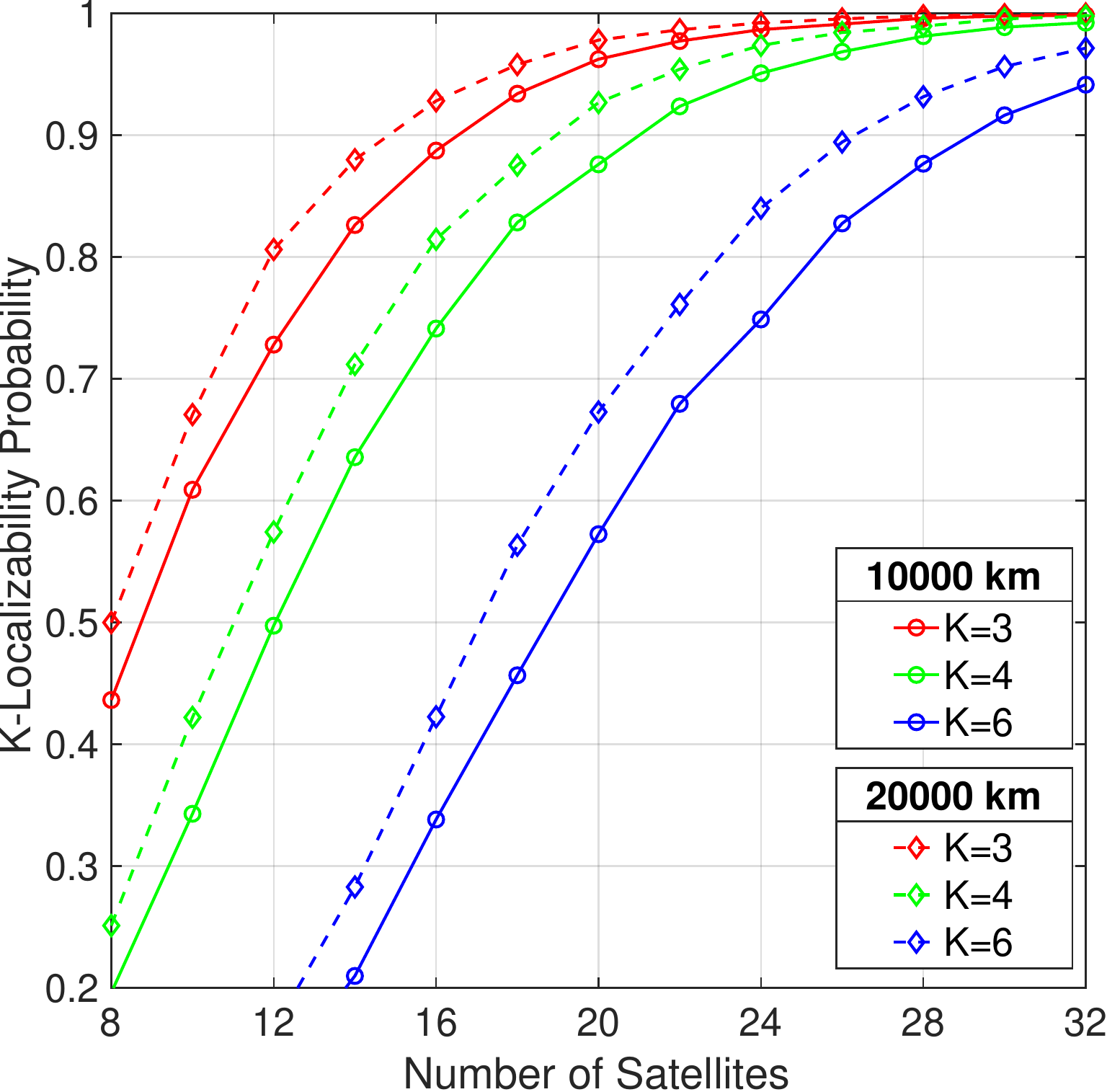}
\caption{Localizability with different constellation configurations.}
\vspace{-0.2cm}
\label{figure5-3}
\end{figure}

The GPS system is operated around an altitude of $20200$~km. The minimum and typical operational MEO satellite numbers for GPS are $24$ and $31$, respectively. When using the time difference of arrival (TDoA) method for localization, at least $4$ satellites are required to participate in the process. As shown in Fig.~\ref{figure5-3}, $24$ MEO satellites with an altitude of $20000$~km can achieve approximately a $97\%$ probability of ensuring at least $4$ satellites can participate in positioning, thus can basically meet the positioning requirements. In addition, the $6-$localizability probability reaches $95\%$ with $31$ MEO satellites. With the same number of satellites, deploying at $20000$~km results in a $5\%-10\%$ increase in localizability probability compared to $10000$~km.

\subsection{Summary and Discussion}
In this subsection, we summarize the common characteristics of these advanced topics.
\begin{itemize}
    \item The SG research based on terrestrial networks has already laid a solid mathematical foundation for these topics. Therefore, the extension of these topics from terrestrial networks to NTN is foreseeable. 
    \item These topics have motivations to be studied within NTN and the spherical SG frameworks. For example, satellite-enabled positioning is considered a mature technology.
    \item Within the spherical SG framework, research on each topic currently involves very little amount of literature. These application scenarios remain underdeveloped, and there are many research directions and contents worth exploring in the future. 
    \item These advanced topics not only involve traditional SG analytical frameworks but also encompass other mathematical tools such as optimization and algorithm design. The content of the research is not limited to communication systems. Compared to terrestrial networks, building analytical frameworks in NTN is more challenging from a mathematical point of view. 
\end{itemize}

\par
{\color{black} Finally, a roadmap table for these advanced topics is provided as Table~\ref{tableV-1}.}

\section{Conclusion}
In this survey, we investigate the modeling and analysis of NTN under the spherical SG framework. As the first survey in the field of spherical SG, it is distinctive, content-comprehensive, and timeline-complete. From unique perspectives such as spherical modeling, three-dimensional topology analysis, and NTN link fading research, this survey is distinctly differentiated from other summary-style literature in its research viewpoints. In addition, this survey comprehensively overviews most of the existing spherical point process models, concepts and derivations in topology, details in aerial and space link channel modeling, and advanced topics with research potential. Furthermore, we address gaps in the current spherical SG framework by incorporating algorithms and case studies, providing technical enhancements in spherical modeling, topology analysis, and advanced topics. Regarding the timeline, we categorize past research into several phases and offer an intuitive review of historical developments. We extensively reference and cover existing studies, conducting a thorough survey of the current state. Discussions on future research directions are integrated into each section, with the final section offering well-reasoned predictions for each advanced topic.

\bibliographystyle{IEEEtran}
\bibliography{references}

% \section*{Biographies}
% \begin{IEEEbiographynophoto}
% {Ruibo Wang} is a Ph.D. student at KAUST, Thuwal, Saudi Arabia. He received his B.S. degree from UESTC in 2020 and his M.S. degree from KAUST in 2022. His current research interests include stochastic geometry, satellite communications, and wireless networks.
% \end{IEEEbiographynophoto}
% \begin{IEEEbiographynophoto}
% {Mohamed-Slim Alouini} [S’94, M’98, SM’03, F’09] received his Ph.D. degree in electrical engineering from the California Institute of Technology, Pasadena, in 1998. He served as a faculty member at the University of Minnesota, Minneapolis, then at Texas A$\&$M University at
% Qatar, Doha, before joining KAUST as a professor of electrical engineering in 2009. His current research interests include the modeling, design, and performance analysis of wireless communication systems.
% \end{IEEEbiographynophoto}

\end{document}